\documentclass[
]{ceurart}

\usepackage{listings}
\usepackage{url}
\usepackage{lineno}
\usepackage{graphicx}
\usepackage{svg}
\usepackage{multirow}
\usepackage{amsmath,amssymb, amsfonts}
\usepackage{amsthm}
\usepackage{mathrsfs}
\usepackage[title]{appendix}
\usepackage{xcolor}
\usepackage{textcomp}
\usepackage{manyfoot}
\usepackage{booktabs}
\usepackage{algorithm}
\usepackage{algorithmicx}
\usepackage{algpseudocode}
\usepackage{listings}
\usepackage{comment}
\usepackage{tikz}
\usepackage{pgfplots}
\usepackage{colortbl}
\pgfplotsset{compat=1.17}
\usepackage{multirow}
\usepackage{csquotes}
\MakeOuterQuote{"}
\usepackage{caption}
\usepackage{threeparttable}
\usepackage{soul}
\usepackage{enumitem}
\usepackage{tipx}
\usepackage{adjustbox}
\usepackage{tcolorbox}
\usepackage{listings}
\usepackage{subcaption}
\usepackage{pdfpages}
\usepackage[normalem]{ulem}
\usepackage{float}

\definecolor{myBlue}{RGB}{70, 130, 180}

\definecolor{myOrange}{RGB}{230, 159, 0}

\definecolor{myGreen}{RGB}{106, 168, 79}
\definecolor{myRed}{RGB}{153, 0, 0}

\usepackage{xspace}
\newcommand{\easer}{EASE$^R$}
\newcommand{\rpb}{$\textrm{RP}^3_{\beta}$}

\newcounter{reqcounter}

\usepackage{etoolbox}
\usepackage{eso-pic}
\usepackage{calc}

\AddToShipoutPictureFG{
  \AtPageUpperLeft{
    \put(\LenToUnit{\dimexpr\paperwidth/2-\textwidth/2\relax}, \LenToUnit{-1.5cm}){
      \vbox{
        \centering
        \begin{minipage}{\textwidth}
          {\small\textbf{Published in the Proceedings of the GENNEXT Workshop at the SIGIR 2025}}\par
          \vspace{2pt}
          \hrule height 0.4pt
        \end{minipage}
      }
    }
  }
}

\begin{document}

\copyrightyear{2025}
\copyrightclause{Copyright for this paper by its authors.
  Use permitted under Creative Commons License Attribution 4.0
  International (CC BY 4.0).}

\conference{GENNEXT@SIGIR'25: The 1st Workshop on Next Generation of IR and Recommender Systems with Language Agents, Generative Models, and Conversational AI, Jul 17, 2025, Padova, Italy}

\title{Exploring Diversity, Novelty, and Popularity Bias in ChatGPT's Recommendations}





\author[1]{Dario Di Palma}[%
email=d.dipalma2@phd.poliba.it,
]
\cormark[1]

\author[1]{Giovanni Maria Biancofiore}[%
email=giovannimaria.biancofiore@poliba.it,
]
\cormark[1]

\author[1]{Vito Walter Anelli}[%
email=vitowalter.anelli@poliba.it,
]

\author[1]{Fedelucio Narducci}[%
email=fedelucio.narducci@poliba.it,
]

\author[1]{Tommaso Di Noia}[%
email=tommaso.dinoia@poliba.it,
]

\address[1]{Politecnico di Bari, Italy}

\cortext[1]{Corresponding author.}

\begin{abstract}
  ChatGPT has emerged as a versatile tool, demonstrating capabilities across diverse domains. Given these successes, the Recommender Systems (RSs) community has begun investigating its applications within recommendation scenarios primarily focusing on accuracy. While the integration of ChatGPT into RSs has garnered significant attention, a comprehensive analysis of its performance across various dimensions remains largely unexplored. Specifically, the capabilities of providing diverse and novel recommendations or exploring potential biases such as popularity bias have not been thoroughly examined. As the use of these models continues to expand, understanding these aspects is crucial for enhancing user satisfaction and achieving long-term personalization. 
  
  This study investigates the recommendations provided by ChatGPT-3.5 and ChatGPT-4 by assessing ChatGPT's capabilities in terms of diversity, novelty, and popularity bias. We evaluate these models on three distinct datasets and assess their performance in Top-N recommendation and cold-start scenarios. The findings reveal that ChatGPT-4 matches or surpasses traditional recommenders, demonstrating the ability to balance novelty and diversity in recommendations. Furthermore, in the cold-start scenario, ChatGPT models exhibit superior performance in both accuracy and novelty, suggesting they can be particularly beneficial for new users. This research highlights the strengths and limitations of ChatGPT's recommendations, offering new perspectives on the capacity of these models to provide recommendations beyond accuracy-focused metrics.
\end{abstract}


\maketitle

\section{Introduction}
Recommender systems (RSs) \cite{DBLP:reference/sp/2022rsh} have long assisted users in discovering valuable information on the web by predicting their preferences and delivering personalized content. Over time, these systems have evolved from Matrix Factorization approaches to modern architectures that extend state-of-the-art Deep Learning models~\cite{DBLP:journals/csur/ZhangYST19}, originally developed for other domains such as time-series forecasting~\cite{DBLP:journals/corr/ChungGCB14,DBLP:journals/corr/abs-2312-00752}, natural language processing~\cite{DBLP:conf/naacl/DevlinCLT19,DBLP:conf/semweb/BellisANS24, servedio2025hidden, di2025llamas}, and computer vision~\cite{AGHILAR2025128579,DBLP:conf/um/AghilarALNRRT25}.
Despite significant progress in improving accuracy, current research in the user modeling and personalization community increasingly emphasizes the importance of beyond-accuracy perspectives such as diversity, novelty, and popularity bias. These factors not only impact overall system effectiveness but also influence user satisfaction~\cite{ping2024beyond}, long-term engagement~\cite{DBLP:journals/fdata/DuricicKLL24}, and fairness~\cite{DBLP:journals/corr/abs-2309-04250}.

With the release of ChatGPT in November 2022
, Large Language Models (LLMs) have begun to reshape how recommendations can be delivered. Unlike traditional RSs that rely on carefully structured training data, LLMs can generate free-form text, potentially offering more nuanced explanations and broader item coverage by leveraging their vast knowledge. Consequently, the research community is now experimenting with LLM-driven recommendation pipelines~\cite{DBLP:conf/recsys/Attimonelli025, DBLP:conf/iir/PalmaSABNCN24, DBLP:conf/ecir/Valentini25}, demonstrating notable successes in improving recommendation accuracy~\cite{DBLP:conf/ecir/HouZLLXMZ24, DBLP:conf/recsys/Palma23, DBLP:conf/recsys/DaiSZYSXS0X23, DBLP:conf/www/AttimonelliDMPG24}. 
However, most existing studies on ChatGPT-based recommender systems have emphasized improving accuracy while neglecting the beyond-accuracy dimensions that are critical for real-world impact \cite{DBLP:conf/recsys/DaiSZYSXS0X23, DBLP:journals/corr/abs-2304-10149}. 

Ignoring the beyond-accuracy behavior of ChatGPT creates a black box for researchers, making it difficult to determine whether it over-recommends popular items, reduces novelty, or offers less diverse recommendations, all of which may negatively impact user satisfaction and long-term personalization goals.
Early investigations focused on how to use ChatGPT for re-ranking recommendations~\cite{10.1145/3700604}, while others began to study the serendipity of the generated recommendations~\cite{DBLP:journals/jaciii/TokutakeO24} or explored how ChatGPT generates recommendations and whether its outputs align more closely with content-based or collaborative filtering approaches~\cite{DBLP:conf/um/PalmaBANN25}.
Only \citet{10.1145/3690655} investigates biases in ChatGPT-based recommender systems, with a specific focus on provider fairness.
While a few works have begun examining potential biases related to sensitive attributes such as race, gender, and religion~\cite{DBLP:conf/recsys/ZhangBZWF023}, aspects like recommendation diversity, novelty, and popularity bias in ChatGPT remain largely unexplored. Addressing these gaps is essential to ensure that personalization technologies are both effective and fair.

To this end, we analyze ChatGPT’s recommendation behavior, focusing on both ChatGPT-3.5 and ChatGPT-4 across multiple beyond-accuracy metrics. Specifically, we investigate whether ChatGPT generates diverse and novel recommendations or exhibits popularity bias, both under normal conditions and in user cold-start scenarios where users have interacted with only a few items. Our evaluation spans three distinct domains, Books, Movies, and Music, using the Facebook Books~\cite{DBLP:conf/recsys/MancinoFBMNS23, DBLP:conf/irongraphs/BufiMFMNS24}, 
MovieLens~\cite{DBLP:journals/tiis/HarperK16}, and Last.FM~\cite{DBLP:conf/recsys/CantadorBK11} datasets as benchmarks, aiming to answer the following Research Questions (RQs):

\begin{enumerate}[label=(RQ\arabic*), itemsep=1pt, parsep=1pt, topsep=1pt, partopsep=1pt]
    \item Are ChatGPT's recommendations diverse?
    \item Are ChatGPT's recommendations novel?
    \item Is ChatGPT affected by popularity bias?
    \item How effective is ChatGPT in user cold-start scenario across accuracy and beyond-accuracy dimensions?
\end{enumerate}



\section{Related Work}\label{sec:related_wor}
\textbf{Diversity, Novelty, and Popularity Bias in Recommender Systems.} Driven by the need for Recommender Systems (RSs) to enhance user engagement~\cite{biancofiore2025conversational}, this work focuses on beyond-accuracy measures of RSs, namely, diversity, novelty, and popularity bias, to investigate how these factors affect the recommendation lists provided by ChatGPT.

There was a moment in the evolution of RSs when researchers realized that evaluating recommendations solely based on accuracy metrics was insufficient. For instance, \citet{DBLP:journals/tois/HerlockerKTR04} suggest that the performance of recommendations should be measured by their usefulness to the user. Similarly \citet{DBLP:journals/mlc/SilveiraZLLM19}, in their survey on the evaluation of RSs, suggest that recommendations can be evaluated based on utility, novelty, diversity, unexpectedness, serendipity, and coverage. Karimi, Jannach, and Jugovac~\cite{DBLP:journals/ipm/KarimiJJ18}, in their review of state-of-the-art news RSs, identify diversity, novelty, and popularity as the most common quality factors for improving recommendations. Specifically, diversity and novelty are often considered quality factors that must be balanced with prediction accuracy~\cite{DBLP:reference/sp/GunawardanaSY22}, and the most discussed beyond-accuracy objectives in recommender system research~\cite{DBLP:journals/tiis/KaminskasB17}. 

As interest in studying RSs beyond accuracy metrics spread, more studies began to use these metrics as goals for improvement. For example, \citet{DBLP:conf/www/ChengWMSX17} and \citet{DBLP:journals/umuai/WuCZ18} focused on creating RSs that not only predict accurate items but also achieve a high level of diversity in recommendations. \citet{DBLP:conf/cikm/NakatsujiFTUFI10} employed a graph-based approach to identify items with higher novelty, while \citet{DBLP:conf/kdd/Cai0WBSWZ024} proposed a method to mitigate popularity bias. Furthermore, the work by \citet{DBLP:conf/recsys/PaparellaPAN23} emphasizes the importance of evaluating RSs beyond accuracy, proposing a multi-objective evaluation approach.

Although many works use beyond-accuracy metrics to evaluate and improve RSs, prior literature lacks a unified framework that rigorously defines diversity, novelty, and bias, leading to vagueness and overlap among these measures. In our study, we define these concepts as follows: Diversity is the extent to which a recommender system suggests a wide range of items from the catalog~\footnote{While we acknowledge that diversity can be measured in multiple ways, such as Top-N diversity (user-level variation in recommendation lists) or temporal diversity (diversity over time), we focus on aggregate diversity due to its measurable implications for item exposure, and long-tail promotion.}, as supported by \cite{DBLP:journals/umuai/JannachLKJ15} and \cite{DBLP:journals/tmis/AdomaviciusZ12}. Novelty is the degree to which recommended items expose users to relevant experiences they are unlikely to discover independently, based on \cite{DBLP:conf/recsys/VargasC11}. Popularity Bias refers to the tendency of recommender systems to favor popular items, those with many interactions, over less popular or niche items, aligning with \cite{DBLP:conf/flairs/AbdollahpouriBM19} and \cite{DBLP:journals/umuai/JannachLKJ15}.\\

\noindent\textbf{ChatGPT-based recommendation.} A first example of ChatGPT for recommendation is proposed by \citet{DBLP:journals/corr/abs-2303-14524}, who introduced ChatREC, a ChatGPT-augmented recommender system that translates the recommendation task into an interactive conversation with users. The authors proposed a prompt template to convert user information and user-item interactions into a query for ChatGPT. However, the system was evaluated solely using accuracy metrics (i.e., Recall, Precision, nDCG). 
\citet{DBLP:conf/um/ManzoorZGJ24} investigates ChatGPT's performance in a multi-turn conversational recommendation setting, demonstrating its potential as a conversational recommender and showing that it outperforms traditional methods. 
\citet{DBLP:conf/ecir/HouZLLXMZ24} focused their work on ChatGPT in zero-shot settings. They analyzed ChatGPT models by designing a dedicated prompting template and revealed that ChatGPT-4 achieved the highest ranking performance compared to other LLMs in the zero-shot recommendation task. 

\citet{DBLP:conf/recsys/SannerBRWD23} investigated the abilities of ChatGPT as a recommender systems for the Top-N recommendation task, aiming to identify the most effective prompting strategy for producing relevant recommendations. The authors concluded that the zero-shot setting yields the most relevant recommendation list, outperforming content-based baselines. However, their conclusions were based solely on nDCG as the evaluation metric, which limits the findings to only one dimension of RSs.

\citet{DBLP:conf/recsys/DaiSZYSXS0X23} investigate ChatGPT's abilities in suggesting items through rating prediction, pairwise recommendation, and re-ranking strategies using the prompting approach. Their experiments, conducted on four domains, demonstrate ChatGPT's abilities to recommend items. Nonetheless, this study provide only an accuracy view of ChatGPTs' capabilities in the Top-N recommendation. 

\citet{li2023bookgpt} focus on applying ChatGPT within the book recommendation scenario, designing BookGPT to address single-item and rating prediction tasks. 
However, the study does not provide a generalizable analysis of ChatGPT's performance across multiple domains, as the authors focus only on the book domain. 


Although all the presented works focus on using ChatGPT to improve the performance of recommender systems, they are primarily based on accuracy metrics. To address this gap, our work investigates the task of Top-N recommendation, moving beyond accuracy by evaluating ChatGPT's performance in terms of diversity, novelty, and popularity bias, while also highlighting its beyond-accuracy capabilities in user cold-start scenarios.

\section{Methodology}\label{sec:methodology}
The following sections discuss the methodology used in our research, outline the design of the prompts employed to collect recommendations from ChatGPT, detail the datasets used in the experiments, present the baselines for comparison, and list the metrics used to assess diversity, novelty, and popularity bias.

\subsection{Prompt Design}\label{sec:prompt_design}
The introduction of GPT-3~\cite{DBLP:conf/nips/BrownMRSKDNSSAA20} demonstrated the ability of LLMs to perform diverse tasks when provided with clear, task-specific prompts, showing how prompts condition the model's response and play a critical role in shaping its performance on a given task~\cite{DBLP:conf/naacl/KongZCLQSZWD24}.

With the widespread diffusion of ChatGPT, the literature on prompt engineering has expanded, moving from basic prompts such as zero- and few-shot~\cite{DBLP:journals/corr/abs-2205-11916} to more complex prompts like Chain-of-Thought~\cite{DBLP:conf/nips/Wei0SBIXCLZ22}, Tree-of-Thoughts~\cite{DBLP:conf/nips/YaoYZS00N23},
Reflexion~\cite{DBLP:conf/nips/ShinnCGNY23}, or Graph-Prompting~\cite{DBLP:conf/www/LiuY0023}. 
Among the various prompt techniques~\cite{DBLP:journals/corr/abs-2402-07927}, we hand-engineered Zero-Shot, Few-Shot, Chain-of-Thought, and Role-Playing (RP) prompting following the works of \citet{DBLP:journals/corr/abs-2401-04997} and 
\citet{li2023bookgpt}, to identify the best approach for our investigation.

In the following, we present the hand-engineered prompts and explain the main reasons for selecting RP prompting as the primary technique for our investigation. Specifically, for all the tested prompts and for each user, the input consists of the user's history, presented as a list of items formatted as follows: $\{History\ of\ the\ User\}: Item_1, Item_2, \ldots, Item_{N}$.

\textbf{Zero-shot prompting}~\cite{radford2019language}\textbf{.} In zero-shot prompting, we directly provided the user's history to ChatGPT and asked for 50 recommendations, as shown in the reference example (see fig.~\ref{fig:zero_shot_prompt}).  However, $\sim$71\% of the generated lists contained fewer than 50 items or included repeated entries, and $\sim$6\% exhibited incorrect task execution.
    
    \begin{figure}[ht]
    \centering
    \adjustbox{center,margin=0.2cm,scale=1}{%
        \begin{tcolorbox}[colback=myOrange!5!white, colframe=myOrange, title=Zero-shot prompt]
            I like \{history of the user\}, provide me 50 recommendations.
        \end{tcolorbox}%
    }
    \caption{Example of a zero-shot prompt designed to generate recommendations based on the user's history.}
    \label{fig:zero_shot_prompt}
\end{figure}

\textbf{Few-shot prompting}~\cite{DBLP:conf/nips/BrownMRSKDNSSAA20}\textbf{.} After zero-shot prompting, we tested few-shot prompting by providing a few demonstrations of recommendations to help the LLM better understand the task (see Fig.~\ref{fig:few_shot_prompt}). While these contextual examples reduced execution errors, $\sim$44\% of the generated lists contained duplicate items.
    
    \begin{figure}[ht]
    \centering
    \adjustbox{center,margin=0.2cm,scale=1}{%
        \begin{tcolorbox}[colback=myOrange!5!white, colframe=myOrange, title=Few-shot prompt]
            User's Watched Movie: "The Shawshank Redemption"\\
            User's Rating: 5 (out of 5)\\
            Recommended Movie: "The Green Mile"\\
            Explanation: Both movies are critically acclaimed drama films with themes of hope and resilience in difficult circumstances. Since the user highly rated "The Shawshank Redemption", they might also enjoy "The Green Mile".\\

            User's Watched Movie: "Inception"\\
            User's Rating: 4\\
            Recommended Movie: "Interstellar"\\
            Explanation: Both movies are science-fiction films directed by Christopher Nolan, known for their mind-bending plots and impressive visuals. A user who enjoyed "Inception" is likely to appreciate "Interstellar" as well.\\

            User's Watched Movie: "The Dark Knight"\\
            User's Rating: 5\\
            Recommended Movie:
        \end{tcolorbox}%
    }
    \caption{Example of a few-shot prompt illustrating recommendations with explanations based on the user's watched movies and ratings.}
    \label{fig:few_shot_prompt}
\end{figure}

\textbf{Chain-of-Thought (CoT) prompting}~\cite{DBLP:conf/nips/Wei0SBIXCLZ22}\textbf{.} Using CoT, we attempted to break the recommendation task into explicit steps to force ChatGPT to reason step-by-step. As shown in Fig.~\ref{fig:chain_of_thought_prompt}, we explicitly defined the instructions, the user's preferences, and the steps to identify the most suitable recommendations. This approach produced excessive tokens, reaching the context limit after generating $\sim$26 items.

    \begin{figure}[ht]
    \centering
    \adjustbox{center,margin=0.2cm,scale=1}{%
        \begin{tcolorbox}[colback=myOrange!5!white, colframe=myOrange, title=Chain-of-Thought prompt]
            Instruction: Recommend a book based on the user's preferences.\\
            
            User's Preferences: The user enjoys science fiction, particularly books with strong character development and intricate world-building. They have previously enjoyed "Dune" by Frank Herbert and "Ender's Game" by Orson Scott Card.\\
            
            Chain of Thought: To recommend a suitable book, I need to consider the user's preferences for science fiction, strong character development, and intricate world-building. The user enjoyed "Dune" and "Ender's Game," which suggests they appreciate complex plots and detailed settings. Based on these criteria, I will identify a book that fits within the science fiction genre and has a reputation for similar qualities.\\
            
            1. The user likes science fiction, so I will focus on books within this genre.\\
            2. The user values strong character development, so I will consider books known for their well-developed characters.\\
            3. The user appreciates intricate world-building, so I will look for books with detailed and immersive settings.\\
            4. Based on their enjoyment of "Dune" and "Ender's Game," I'll look for books with complex plots and critical acclaim.\\
            
            Recommendation: "The Left Hand of Darkness" by Ursula K. Le Guin\\
            
            Explanation: "The Left Hand of Darkness" is a renowned science fiction novel known for its detailed world-building and exploration of complex social and cultural issues. Ursula K. Le Guin's character development is exceptional, and the story's setting on the planet Gethen provides a rich and immersive experience. This book should align well with the user's preferences and previous enjoyment of "Dune" and "Ender's Game."
        \end{tcolorbox}%
    }
    \caption{Example of a Chain-of-Thought prompt for book recommendation, incorporating user preferences and step-by-step reasoning to arrive at a recommendation.}
    \label{fig:chain_of_thought_prompt}
\end{figure}

\textbf{Role-Playing prompting}~\cite{DBLP:conf/nips/JinC0YF00W23}\textbf{.} Following the work of \citet{DBLP:journals/corr/abs-2401-04997} and \citet{li2023bookgpt}, we also tested Role-Playing prompts, where ChatGPT impersonates a Recommender System and recommends items based on the user's history (see Fig.~\ref{fig:top_n_rec_prompt}). This strategy
proved the most effective, eliminating duplicate recommendations.

    \begin{figure}[ht]
    \centering
    \adjustbox{center,margin=0.2cm,scale=1}{%
        \begin{tcolorbox}[colback=myBlue!5!white, colframe=myBlue!92!black, title=Role-Playing Recommender Prompt]
            Given a user, as a Recommender System, please provide only the names of the top 50 recommendations.\\
            You know that the user likes the following items: \{history of the user\}
        \end{tcolorbox}%
    }
    \caption{Example of a Role-Playing Recommender prompt designed to generate a ranked list of 50 recommendations based on the user's history.}
    \label{fig:top_n_rec_prompt}
\end{figure}

After testing 30 hand-crafted prompts and aligning with studies on Role-Playing Prompting~\cite{DBLP:journals/corr/abs-2401-04997, DBLP:journals/corr/abs-2308-07702}, we selected this approach for its ability to reduce duplicates and token usage. In this setup, ChatGPT acts as a Recommender System, generating 50 recommendations based on the user’s history (see Fig.~\ref{fig:top_n_rec_prompt}).


\subsection{Experimental Setup}\label{sec:exp_setup}
This section outlines the experimental setup, including the datasets, baselines, and metrics used to assess the beyond-accuracy performance of ChatGPT's recommendations, with a focus on diversity, novelty, and popularity bias. 

\noindent\textbf{Datasets.} We evaluated ChatGPT on three well-known recommendation datasets, namely MovieLens100k~\cite{DBLP:journals/tiis/HarperK16}, Last.FM~\cite{DBLP:conf/recsys/CantadorBK11}, and Facebook Books~\footnote{\url{https://2015.eswc-conferences.org/program/semwebeval.html}}. To enhance data quality, we applied an iterative 10-core filtering strategy~\cite{DBLP:conf/sigir/MancinoA25}, retaining only users and items with at least ten interactions. Table~\ref{tab:dataset_statistics} holds the dataset statistics after preprocessing.
\begin{table}[t!]
\centering
\caption{Dataset statistics after pre-processing with \scalebox{0.9}{$k-core \geq\ 10$}.}
\label{tab:dataset_statistics}

\begin{tabular}{lccccc}
\toprule
\textbf{Dataset} & \textbf{Interaction} & \textbf{Users} & \textbf{Items} & \textbf{Sparsity} & \textbf{Content} \\
\midrule
\textbf{MovieLens} & 42,456 & 603 & 1,862 & 96.22\% & genre \\
\textbf{Last.FM} & 49,171 & 1,797 & 1,507 & 98.18\% & genre \\
\textbf{FB Books} & 13,117 & 1,398 & 2,234 & 99.58\% & genre, author \\
\bottomrule
\end{tabular}
\end{table}

\begin{table}[t!]
\centering
\caption{Overview of beyond-accuracy metrics}
\label{tab:beyond_accuracy_table}


\begin{tabular}{@{}p{0.25\textwidth}p{0.15\textwidth}p{0.55\textwidth}@{}}
\toprule
\textbf{Aspect} & \textbf{Metric} & \textbf{Description} \\
\midrule
\multirow{2}{=}{Diversity} & ItemCV & 
Item Coverage (ItemCV) measures how many items appear in the top-$n$ recommendations of users, ensuring a broader selection of content is provided. \\
\cmidrule{2-3}
& Gini & Gini Index: A measure of statistical dispersion intended to represent the inequality of a distribution. The Gini Index ranges between 0 and 1, where a higher value indicates greater concentration of recommendations, e.g., on popular items~\cite{DBLP:journals/umuai/JannachLKJ15}. We report \(1-\text{Gini}\) where higher values indicate less concentrated recommendations. \\
\midrule
\multirow{2}{=}{Novelty} & EFD & Expected Free Discovery (EFD): A novelty measure based on the inverse collection frequency, expressing the algorithm's ability to recommend relevant long-tail items. Recommending such items introduces users to less obvious, unique content, enriching the user experience~\cite{DBLP:conf/recsys/VargasC11}. \\
\cmidrule{2-3}
& EPC & Expected Popularity Complement (EPC): This metric quantifies the "number of unseen items now seen," promoting the discovery of previously unknown content and supporting user exploration~\cite{DBLP:conf/recsys/VargasC11}. \\
\midrule
\multirow{2}{=}{Popularity Bias} & APLT & Average Popularity of Long-Tail Items (APLT): Measures the average popularity of long-tail items in the top-$n$ recommendations, ensuring less mainstream items are highlighted~\cite{DBLP:conf/flairs/AbdollahpouriBM19, DBLP:conf/cikm/PaparellaAN0N23}. \\
\cmidrule{2-3}
& ARP & Average Rating-based Popularity (ARP): Computes the popularity of items in a recommendation list based on the number of interactions each item has in the training data. By considering item popularity, this metric helps balance recommendations to avoid overexposure of popular items, addressing biases and ensuring a more equitable distribution for varied user preferences~\cite{DBLP:journals/umuai/JannachLKJ15}. \\
\bottomrule
\end{tabular}

\end{table}

\noindent\textbf{Baseline Models.} To measure the effectiveness of ChatGPT, we experimentally compare its performance with state-of-the-art baselines from three categories: Non-Personalized, Collaborative Filtering, and Content-Based Filtering methods. To ensure a fair comparison, we train the baselines and optimize their hyperparameters using the Elliot framework~\cite{DBLP:conf/sigir/AnelliBFMMPDN21}, and split the dataset into 80\% training and 20\% test sets, following the all unrated items  evaluation protocol ~\cite{10.1145/3588901,javed2021review}. The code used for the experiments is publicly available at: \url{https://github.com/sisinflab/beyond-accuracy-recsys-chatgpt}. Below, we describe the baselines, grouped by recommendation category. \textit{Non-Personalized.} Random and Most Popular return random recommendations and the most popular recommendations, respectively, and are used as reference points. \textit{Collaborative Filtering.} To compare the effect of ChatGPT recommendations on beyond-accuracy metrics, we selected the following collaborative filtering methods, each focusing on different aspects. Specifically, we selected \rpb~\cite{DBLP:journals/tiis/PaudelCNB17} and LightGCN~\cite{DBLP:conf/sigir/0001DWLZ020} for their demonstrated ability to maintain accuracy while preserving diversity~\cite{DBLP:conf/www/CooperLRS14, DBLP:conf/sigir/0001DWLZ020}. ItemKNN~\cite{DBLP:conf/sigecom/SarwarKKR00}, UserKNN~\cite{DBLP:conf/uai/BreeseHK98}, and \easer~\cite{DBLP:conf/www/Steck19} were chosen for their emphasis on relevance and personalization~\cite{DBLP:conf/www/Steck19, DBLP:conf/sigecom/SarwarKKR00}. Finally, MF2020~\cite{DBLP:conf/recsys/RendleKZA20} and NeuMF~\cite{DBLP:conf/www/HeLZNHC17} were included as a tradeoff between model complexity and effectiveness.
\textit{Content-Based Filtering.} We further extend our comparison by including content-based models, which prioritize explicit feature representations and offer a meaningful contrast to collaborative models. This allows us to evaluate ChatGPT against the most appropriate model type for the dataset. Specifically, we include VSM~\cite{DBLP:journals/cacm/SaltonWY75}
which represents items as vectors in a high-dimensional space, with each dimension corresponding to a feature, as well as AttributeItemKNN and AttributeUserKNN~\cite{DBLP:conf/recsys/GantnerRFS11}, which rely on TF-IDF-weighted attribute vectors to compute similarities and generate recommendations.\\
\noindent \textbf{Ensuring Recommendation Consistency.} ChatGPT models generate recommendations based solely on the user profile provided in the prompt, without being constrained to a predefined dataset. As a result, they may hallucinate~\cite{DBLP:conf/cikm/ChenFYWFL0LX23, DBLP:conf/acl/NieYWPL19, DBLP:journals/csur/JiLFYSXIBMF23} or suggest real items not present in the reference dataset, leading to discrepancies in item names and inconsistencies in evaluation.

To address this, we adopt a post-processing pipeline that uses Gestalt pattern matching~\cite{DBLP:journals/iandc/GiulianoJKMS61} to identify the closest match in the dataset, accepting items with a similarity score above 90\% (empirically determined). Unmatched items are flagged as External Items, originating from the LLM’s pre-trained knowledge, and excluded from evaluation to ensure a fair comparison with traditional recommenders by selecting in-catalogue items.

Since this final step could affect our evaluations, we verified that out-of-catalogue items consistently appeared beyond the top-10 positions in all recommendation lists, ensuring that rank-sensitive metrics remain unaffected and preserving the validity of our evaluation. In our configuration, ChatGPT placed these items only after rank 23, suggesting 2,740 out-of-catalogue items for Books, 870 for Music, and 234 for Movies.

Finally, to ensure a fair comparison, we evaluate all models and ChatGPT results at a cutoff of 10 (i.e., Top-10 recommendations per user), following widely accepted practices in recommendation~\cite{DBLP:conf/recsys/PetrovM23, DBLP:conf/recsys/RendleKZA20, DBLP:conf/www/HeLZNHC17}.\\
\noindent\textbf{Evaluation Metrics.} While our primary focus is on the beyond-accuracy aspects of ChatGPT's recommendations, it is also important to include \textit{accuracy metrics} to assess whether the recommendations are relevant to users. For this purpose, we use two standard metrics: Precision and Recall~\cite{DBLP:reference/sp/2022rsh, DBLP:books/daglib/0022145}. Higher values of Precision and Recall indicate that the recommender system provides a greater number of relevant items. Additionally, we evaluate the ranking quality of the recommendations using the Normalized Discounted Cumulative Gain (nDCG)~\cite{DBLP:journals/tois/JarvelinK02}, where higher values indicate better recommendation lists.

For \textit{beyond-accuracy metrics}, we selected a set of measures to evaluate diversity, novelty, and popularity bias. The specific metrics considered are detailed in Table~\ref{tab:beyond_accuracy_table}.
\section{Experimental Results}\label{sec:exp_results}
\subsection{ChatGPT Beyond-Accuracy Recommendation Performance}\label{sec:classical_rec_performance}
In this section, we discuss the empirical findings from Table~\ref{tab:combined_all}, focusing on (RQ1.) the diversity of ChatGPT’s recommendations, (RQ2.) their novelty, and (RQ3.) the extent to which ChatGPT is affected by popularity bias. The evaluation comprises three datasets, Facebook Books, Last.FM, and MovieLens, and compares ChatGPT‑3.5 and ChatGPT‑4 against both Collaborative Filtering and Content‑Based Filtering baselines. Statistically significant differences (paired t‑tests at $p<0.05$) are noted where indicated in the table.\\

\noindent\textbf{Preliminary Accuracy Analysis.}
Before examining diversity, novelty, and popularity bias, we first verify that ChatGPT’s recommendations fulfill the primary goal of offering relevant items. We use nDCG, Recall, and Precision as standard accuracy metrics. Higher values on these metrics imply better recommendation.

Overall, ChatGPT demonstrates a comparable level of accuracy in recommendation scenarios. Specifically, on \underline{Facebook Books}, ChatGPT-4 attains the highest nDCG overall (0.0932), significantly outperforming the best baseline, AttributeItemKNN (0.0479), as well as ChatGPT-3.5 (0.0668). Recall and Precision follow a similar pattern to nDCG.

For \underline{Last.FM}, while ChatGPT-4 (nDCG = 0.2832) does not surpass the best Collaborative Filtering (CF) approach (\rpb: 0.3147), it still ranks among the top-performing algorithms. ChatGPT-3.5 trails behind ChatGPT-4 but still outperforms some baselines (e.g., \easer, AttributeItemKNN).

For \underline{MovieLens}, although ChatGPT-4 improves upon ChatGPT-3.5 across all accuracy metrics, raising nDCG from 0.1475 to 0.1815 and Precision from 0.1120 to 0.1551, certain CF algorithms (e.g., \rpb: 0.2827 nDCG, 0.2708 Precision) achieve significantly higher scores. Nonetheless, ChatGPT’s accuracy levels comfortably exceed those of some methods, such as VSM (0.0174 nDCG) and AttributeItemKNN (0.0326 nDCG).

These results demonstrate that both ChatGPT-3.5 and ChatGPT-4 achieve valid and reasonable performance on accuracy metrics. This preliminary evaluation ensures that the subsequent analysis of diversity, novelty, and popularity bias is based on recommendations that already meet the accuracy standard. In the following sections, our analysis is divided according to the research questions (RQs).\\

\begin{table*}[t!]
\centering
\caption{\small{Combined results across three datasets (Facebook Books, Last.FM, and MovieLens). Preferred metric values are indicated by arrows (\textuparrow for higher, \textdownarrow for lower). Best values are in bold, and second-best are underlined. Results are ranked by nDCG. Baseline results are statistically significant (paired t-tests, $p < 0.05$) unless marked with *(ChatGPT-3.5) or \textdagger(ChatGPT-4). `Best-CF' and `CBF' denote the top Collaborative Filtering and Content-Based Filtering baselines (by nDCG) for each dataset.}}
\label{tab:combined_all}
\vspace{3pt}
\resizebox{0.9\textwidth}{!}{%
\begin{tabular}{lccccccccc}
\toprule
\multicolumn{10}{c}{\textbf{Facebook Books}} \\
\midrule
\textbf{Model} 
& \multicolumn{3}{c}{\textbf{Accuracy}} 
& \multicolumn{2}{c}{\textbf{Diversity}} 
& \multicolumn{2}{c}{\textbf{Novelty}} 
& \multicolumn{2}{c}{\textbf{Popularity Bias}} \\
\cmidrule(lr){2-4}\cmidrule(lr){5-6}\cmidrule(lr){7-8}\cmidrule(lr){9-10}
& \textbf{nDCG} \textuparrow & \textbf{Recall} \textuparrow & \textbf{Precision} \textuparrow & \textbf{Gini} \textuparrow & \textbf{ItemCV} \textuparrow & \textbf{EPC} \textuparrow & \textbf{EFD} \textuparrow & \textbf{APLT} \textuparrow & \textbf{ARP} \textdownarrow \\
\midrule
Random                                & 0.0019 & 0.0034 & 0.0008 & 0.7753 & 2230 & 0.0007 & 0.0078 & 0.6874 & 5.7186 \\
MostPop                               & 0.0091 & 0.0137 & 0.0033 & 0.0045 & 17   & 0.0031 & 0.0228 & 0.0000 & 138.3632 \\
LightGCN                              & 0.0105 & 0.0171 & 0.0038 & 0.0053 & 112  & 0.0035 & 0.0269 & 0.0132 & 134.0763 \\
NeuMF                                 & 0.0167 & 0.0243 & 0.0057 & \underline{0.3336} & 1563 & 0.0065 & 0.0661 & 0.2444\textdagger & 16.0072 \\
\easer                                & 0.0188 & 0.0313 & 0.0071 & 0.0111 & 228  & 0.0066 & 0.0547 & 0.0032 & 125.2026 \\
ItemKNN                               & 0.0288 & 0.0408 & 0.0086 & \textbf{0.5293} & \textbf{2141} & 0.0104 & 0.1099 & \textbf{0.5974} & 24.9652 \\
MF2020                                & 0.0317 & 0.0592 & 0.0133 & 0.0044 & 15   & 0.0116 & 0.0953 & 0.0000 & 114.0167 \\
UserKNN                               & 0.0320 & 0.0468 & 0.0098 & 0.1564 & 1372 & 0.0115 & 0.1065 & 0.0852 & 55.2988 \\
\rpb\ \textsuperscript{\textbf{BestCF}}   & 0.0379 & 0.0568 & 0.0120 & 0.3063 & \underline{1888} & 0.0138 & 0.1357 & 0.3308 & 44.1225* \\
AttributeUserKNN                      & 0.0402 & 0.0593 & 0.0133 & 0.0918 & 945  & 0.0152 & 0.1414 & 0.0466 & 64.2887 \\
VSM                                   & 0.0458 & 0.0785 & 0.0172 & 0.2478 & 1389 & 0.0173 & 0.1913 & 0.5761 & \underline{7.3705} \\
AttributeItemKNN\ \textsuperscript{\textbf{BestCBF}} 
                                      & 0.0479 & 0.0705 & 0.0155 & 0.2824 & 1510 & 0.0182 & 0.2019 & \underline{0.5879} & \textbf{7.1801} \\
\cmidrule(lr){1-10}
ChatGPT-3.5                           & \underline{0.0668} & \underline{0.0936} & \underline{0.0205} & 0.0713 & 853  & \underline{0.0250} & \underline{0.2480} & 0.1870 & 46.3236 \\
ChatGPT-4                             & \textbf{0.0932}    & \textbf{0.1283}    & \textbf{0.0283}    & 0.1050 & 1004 & \textbf{0.0353}    & \textbf{0.3486}    & 0.2424 & 40.0319 \\
\bottomrule
\addlinespace[2pt] 
\toprule
\multicolumn{10}{c}{\textbf{Last.FM}} \\
\midrule
\textbf{Model} 
& \multicolumn{3}{c}{\textbf{Accuracy}} 
& \multicolumn{2}{c}{\textbf{Diversity}} 
& \multicolumn{2}{c}{\textbf{Novelty}} 
& \multicolumn{2}{c}{\textbf{Popularity Bias}} \\
\cmidrule(lr){2-4}\cmidrule(lr){5-6}\cmidrule(lr){7-8}\cmidrule(lr){9-10}
& \textbf{nDCG} \textuparrow & \textbf{Recall} \textuparrow & \textbf{Precision} \textuparrow & \textbf{Gini} \textuparrow & \textbf{ItemCV} \textuparrow & \textbf{EPC} \textuparrow & \textbf{EFD} \textuparrow & \textbf{APLT} \textuparrow & \textbf{ARP} \textdownarrow \\
\midrule
Random                                         & 0.0044          & 0.0068          & 0.0052          & 0.8398          & 1507          & 0.0045          & 0.0478          & 0.5678          & 31.6844 \\
NeuMF                                          & 0.1005          & 0.1133          & 0.0860          & \textbf{0.5049} & \textbf{1492} & 0.0804          & 0.7848          & \underline{0.2418}    & \textbf{77.5480} \\
MostPop                                        & 0.1009          & 0.0895          & 0.0740          & 0.0081          & 27            & 0.0662          & 0.5907          & 0.0000          & 348.3308 \\
LightGCN                                       & 0.1408          & 0.1329          & 0.1060          & 0.1114          & 635           & 0.1013          & 0.9372          & 0.2063          & 135.0381 \\
AttributeItemKNN                               & 0.2233          & 0.2013*         & 0.1481*         & \underline{0.3854}     & \underline{1411}     & 0.1584          & 1.5710          & \textbf{0.3043} & \underline{87.8647} \\
\easer                                         & 0.2278          & 0.1949*         & 0.1509          & 0.0331          & 283           & 0.1517          & 1.3761          & 0.0088          & 247.6099 \\
VSM                                            & 0.2451*         & 0.2021*         & 0.1511          & 0.0826          & 653           & 0.1593          & 1.4845          & 0.0585          & 177.9949 \\
AttributeUserKNN\ \textsuperscript{\textbf{BestCBF}} 
                                               & 0.2795\textdagger & 0.2364          & 0.1818\textdagger & 0.1724          & 923           & 0.1947\textdagger & 1.8297\textdagger & 0.0895          & 134.5766 \\
UserKNN                                        & 0.2983          & 0.2538          & 0.1912          & 0.1491          & 846           & 0.2030          & 1.9060\textdagger & 0.0550          & 152.7412 \\
ItemKNN                                        & 0.3013          & \underline{0.2595}     & 0.1925          & 0.1634          & 962           & 0.2080          & \underline{1.9854}     & 0.1146          & 152.4739 \\
MF2020                                         & \underline{0.3097}     & 0.2576          & \textbf{0.1986} & 0.0908          & 460           & \textbf{0.2116} & 1.9571          & 0.0051          & 181.8922 \\
\rpb\ \textsuperscript{\textbf{BestCF}}       & \textbf{0.3147} & \textbf{0.2634} & \underline{0.1957}     & 0.1441          & 831           & \underline{0.2110} & \textbf{1.9970} & 0.0678          & 153.0884 \\
\cmidrule(lr){1-10}
ChatGPT-3.5                                    & 0.2448          & 0.1964          & 0.1408          & 0.1927          & 952           & 0.1680          & 1.6436          & 0.1391          & 99.3311 \\
ChatGPT-4                                      & 0.2832          & 0.2313          & 0.1680          & 0.2023          & 944           & 0.1918          & 1.8663          & 0.1267          & 102.1045 \\
\bottomrule
\addlinespace[2pt] 
\toprule
\multicolumn{10}{c}{\textbf{MovieLens}} \\
\midrule
\textbf{Model} 
& \multicolumn{3}{c}{\textbf{Accuracy}} 
& \multicolumn{2}{c}{\textbf{Diversity}} 
& \multicolumn{2}{c}{\textbf{Novelty}} 
& \multicolumn{2}{c}{\textbf{Popularity Bias}} \\
\cmidrule(lr){2-4}\cmidrule(lr){5-6}\cmidrule(lr){7-8}\cmidrule(lr){9-10}
& \textbf{nDCG} \textuparrow & \textbf{Recall} \textuparrow & \textbf{Precision} \textuparrow & \textbf{Gini} \textuparrow & \textbf{ItemCV} \textuparrow & \textbf{EPC} \textuparrow & \textbf{EFD} \textuparrow & \textbf{APLT} \textuparrow & \textbf{ARP} \textdownarrow \\
\midrule
Random                                & 0.0087 & 0.0062 & 0.0129 & 0.6917          & 1776          & 0.0108          & 0.1230          & 0.5482          & 22.2227 \\
VSM                                   & 0.0174 & 0.0099 & 0.0205 & 0.0529          & 409           & 0.0209          & 0.2305          & \underline{0.3732}     & \underline{29.2857} \\
AttributeItemKNN                      & 0.0326 & 0.0220 & 0.0389 & \textbf{0.3962} & \textbf{1395} & 0.0375          & 0.4285          & \textbf{0.5510} & \textbf{23.6326} \\
LightGCN                              & 0.0411 & 0.0349 & 0.0500 & \underline{0.3105}     & 1136          & 0.0421          & 0.4637          & 0.3040          & 43.4723 \\
NeuMF                                 & 0.1235 & 0.0999\textdagger & 0.1324 & 0.2761          & \underline{1172} & 0.1171*         & 1.2757*         & 0.0970          & 70.5342 \\
MostPop                               & 0.1488* & 0.0841* & 0.1424\textdagger & 0.0083          & 40            & 0.1097          & 1.2750*         & 0.0000          & 182.4909 \\
MF2020                                & 0.2013 & 0.1298 & 0.1985 & 0.0173          & 94            & 0.1576\textdagger & 1.7712          & 0.0000          & 162.5163 \\
\easer                                & 0.2076 & 0.1229 & 0.1872 & 0.0118          & 67            & 0.1522\textdagger & 1.7352\textdagger & 0.0000          & 173.2040 \\
AttributeUserKNN\ \textsuperscript{\textbf{BestCBF}} 
                                      & 0.2152 & 0.1317 & 0.2045 & 0.0590 & 438 & 0.1743 & 1.9356 & 0.0117 & 127.1064 \\
ItemKNN                               & 0.2709 & \underline{0.1819} & \underline{0.2626} & 0.1036 & 666 & \underline{0.2348} & \underline{2.5547} & 0.0470 & 103.0248 \\
UserKNN                               & \underline{0.2814} & 0.1769 & 0.2601 & 0.0589 & 428 & 0.2263 & 2.4958 & 0.0057 & 125.7174 \\
\rpb\ \textsuperscript{\textbf{BestCF}} 
                                      & \textbf{0.2827} & \textbf{0.1898} & \textbf{0.2708} & 0.1230 & 744 & \textbf{0.2421} & \textbf{2.6613} & 0.0643*\textdagger & 100.4106 \\
\cmidrule(lr){1-10}
ChatGPT-3.5                           & 0.1475 & 0.0807 & 0.1120 & 0.0851 & 591 & 0.1260 & 1.3981 & 0.0733\textdagger & 90.7590 \\
ChatGPT-4                             & 0.1815 & 0.1109 & 0.1551 & 0.0853 & 553 & 0.1453 & 1.6010 & 0.0775* & 95.7042 \\
\bottomrule
\end{tabular}
} 
\end{table*}

\noindent\large{\textbf{(RQ1.) Are ChatGPT's recommendations diverse?}}
We assess diversity using Gini and Item Coverage (ItemCV). A lower Gini indicates a higher concentration toward certain items, while higher coverage values indicate that more items from the catalog are recommended.

\textbf{Facebook Books (Table~\ref{tab:combined_all}).} ChatGPT-4 achieves a Gini of 0.1050 and an ItemCV of 1,004, outperforming ChatGPT-3.5 (Gini = 0.0713, ItemCV = 853) on both metrics. Although several baselines, such as ItemKNN (Gini = 0.5293, ItemCV = 2,141), still yield a better diversity score, both ChatGPT-4 and ChatGPT-3.5 generally rank above baselines such as MostPop and \easer. In terms of item coverage and Gini, ChatGPT models demonstrate a high concentration on specific items while covering nearly half of the total span (1,004 out of 2,234 items).

\textbf{Last.FM (Table~\ref{tab:combined_all}).} A similar trend emerges: ChatGPT-4 has a higher Gini (0.2023) than ChatGPT-3.5 (0.1927), indicating a lower concentration of recommendations on specific items. Additionally, GPT-4 covers 944 out of 1,507 items, whereas \rpb, which is designed to trade off diversity and accuracy, achieves a coverage value of 831 and a Gini of 0.1441, demonstrating ChatGPT's strong ability to recommend diverse items.

\textbf{MovieLens (Table~\ref{tab:combined_all}).} ChatGPT-4 achieves a Gini of 0.0853, a slight improvement over ChatGPT-3.5 (0.0851). However, its item coverage spans 553 out of 1,862 items, which is comparatively lower than approaches such as \rpb (Gini = 0.1230, ItemCV = 744). These results highlight that, although the diversity score is lower than certain baselines, ChatGPT still presents a comparable diversity score on this dataset.\\

\noindent\underline{Summary (RQ1).} \textit{ChatGPT’s recommendations are moderately diverse for Facebook Books and Last.FM, while exhibiting limited diversity on MovieLens, with GPT‑4 consistently outperforming GPT‑3.5. Although it does not match the highest-diversity baselines, it shows superior diversity compared to some CF and CBF approaches.}\\

\noindent\large{\textbf{(RQ2.) Are ChatGPT's recommendations novel?}} Novelty is measured via EPC (Expected Popularity Complement) and EFD (Expected Free Discovery), both interpreted such that higher values imply more novel recommendations.

\textbf{Facebook Books (Table~\ref{tab:combined_all}).} ChatGPT‑4 exhibits relatively high novelty (EPC=0.0353, EFD=0.3486), exceeding most baselines, including ChatGPT‑3.5 (EPC=0.0250, EFD=0.2480), and even surpassing all CF and CBF algorithms on these metrics.

\textbf{Last.FM (Table~\ref{tab:combined_all}).} Both ChatGPT versions rank above average in EPC and EFD, with CF and CBF methods (e.g., \rpb: EPC=0.2110, EFD=1.9970, VSM: EPC=0.1593, EFD=1.4845) performing at a comparable level. Still, the difference between ChatGPT‑4 (0.1918 EPC, 1.8663 EFD) and ChatGPT‑3.5 (0.1680 EPC, 1.6436 EFD) suggests GPT‑4 more effectively recommends less mainstream items.

\textbf{MovieLens (Table~\ref{tab:combined_all}).}
On MovieLens, ChatGPT-4 (0.1453 EPC, 1.6010 EFD) outperforms ChatGPT-3.5 (0.1260 EPC, 1.3981 EFD) in terms of EPC and EFD values and places it on par with other methods (e.g., NeuMF: 0.1171 EPC, 1.2767 EFD), although lower than \rpb (EPC of 0.2421, EFD of 2.6613), the best model.\\

\noindent\underline{Summary (RQ2).} \textit{ChatGPT’s recommendations exhibit above-average novelty in MovieLens and high novelty in Facebook Books and Last.FM, with GPT-4 generally surpassing GPT-3.5. The results suggest that ChatGPT, based on the user's history, also recommends novel items for each user.}\\

\noindent\large{\textbf{(RQ3.) Is ChatGPT affected by popularity bias?}}
We examine popularity bias using APLT (Average Popularity of Long-Tail Items; higher indicates stronger inclination toward long-tail (less popular) items) and ARP (Average Rating-based Popularity; lower values imply less popularity bias).

\textbf{Facebook Books (Table~\ref{tab:combined_all}).} ChatGPT-3.5's recommendations yield APLT = 0.1870 and ARP = 46, while ChatGPT-4 improves to APLT = 0.2424 and ARP = 40. With a higher APLT and lower ARP, GPT-4 demonstrates a better capability for recommending long-tail and less popular items than GPT-3.5. Although both models remain far from pure MostPop methods (ARP = 138), some baselines, such as AttributeItemKNN (APLT = 0.5879, ARP = 7) and VSM (APLT = 0.5761, ARP = 7), achieve better APLT and ARP values.

\textbf{Last.FM (Table~\ref{tab:combined_all}).} ChatGPT-3.5 has an APLT of 0.1391 and an ARP of 99, while GPT-4 has an APLT of 0.1267 and an ARP of 102, positioning it in the mid-range of models. This suggests that GPT-4 covers a smaller percentage of the long tail and tends to recommend more popular items. Although it outperforms certain baselines, such as \rpb (APLT = 0.0678, ARP = 153), it does not perform as well as other baselines, such as AttributeItemKNN (APLT = 0.3043, ARP = 87.8647).

\textbf{MovieLens (Table~\ref{tab:combined_all}).} ChatGPT shows an ARP of 90 for GPT-3.5 and 95 for GPT-4, which is lower than MostPop (ARP = 182) but higher than some graph-based methods (e.g., LightGCN: ARP = 43) or neighbor-based methods (e.g., AttributeItemKNN: ARP = 23). This indicates that its behavior is not as popularity-driven as MostPop but is still influenced by popular items. A similar trend is observed for APLT, further demonstrating that ChatGPT does not recommend items from the long tail and exhibits a degree of popularity bias.\\

\noindent\underline{Summary (RQ3).} \textit{Although ChatGPT's values are far from those obtained by MostPop, it still exhibits a tendency to recommend popular items, neglecting items in the long tail. In particular, GPT-4 demonstrates a lower ARP than ChatGPT-3.5, suggesting a tendency to recommend less popular items.}\\

\noindent\textit{\textbf{To conclude,}} ChatGPT models exhibit strong beyond-accuracy performance, achieving an optimal balance of novelty and diversity in the books domain, comparable results in the music domain, and suboptimal outcomes in the movie domain. Although it shows some inclination toward popular items, this bias is far less pronounced compared to MostPop or other strongly popularity-biased baselines. Furthermore, the improvements observed from GPT-3.5 to GPT-4 across all three datasets highlight the strength of GPT-4 for recommendations, particularly in balancing beyond-accuracy trade-offs.

These findings underscore the potential of ChatGPT as a recommender system while also highlighting areas for improvement, particularly in refining its ability to balance relevance, diversity, and novelty across domains.

\begin{table*}[t!]
\centering
\caption{\small{Comparative Analysis of \textbf{User Cold Start} Interactions (Maximum of Ten Interactions per User) with ChatGPT-3.5 and ChatGPT-4. Arrows indicate whether higher (\textuparrow) or lower (\textdownarrow) values are desirable for each metric. Best values are in bold, and second-best values are underlined. CF and CBF represent Collaborative Filtering and Content-based Filtering recommenders. The Facebook Books baselines are statistically significant based on paired t-tests ($p < 0.05$) except for the values denoted with *(for ChatGPT-3.5) and \dag(for ChatGPT-4). Best-CF and CBF correspond to the best Collaborative Filtering and Content-Based Filtering based on the nDCG.}}

\label{tab:coldstart_10}

\resizebox{0.9\textwidth}{!}{ 

\begin{tabular}{clcccccccccc}

\toprule
& \textbf{Model} & \multicolumn{3}{c}{\textbf{Accuracy}} & \multicolumn{2}{c}{\textbf{Diversity}} & \multicolumn{2}{c}{\textbf{Novelty}} & \multicolumn{2}{c}{\textbf{Popularity Bias}} \\
\cmidrule(lr){3-5} \cmidrule(lr){6-7} \cmidrule(lr){8-9} \cmidrule(lr){10-11}
& & \textbf{nDCG} \textuparrow & \textbf{Recall} \textuparrow & \textbf{Precision} \textuparrow & \textbf{Gini} \textuparrow & \textbf{ItemCV} \textuparrow & \textbf{EPC} \textuparrow & \textbf{EFD} \textuparrow & \textbf{APLT} \textuparrow & \textbf{ARP} \textdownarrow \\

\midrule

\multirow{6}{*}{\rotatebox[origin=c]{90}{\textbf{Facebook Books}}}
& Random           & 0.0011         & 0.0018         & 0.0004         & 0.4315         & 1560         & 0.0004         & 0.0034\dag     & 0.6793         & 5.8468          \\
& MostPop          & 0.0115*        & 0.0143         & 0.0029         & 0.0037         & 14           & 0.0031         & 0.0236         & 0.0000         & 139.1043        \\
& AttributeItemKNN \textsuperscript{\textbf{CBF}} & 0.0335*    & 0.0500         & 0.0100*        & \ul{0.1873}   & \ul{957}       & 0.0119*        & 0.1283*        & \textbf{0.5661} & \textbf{7.2114} \\
& \rpb \textsuperscript{\textbf{CF}}          & 0.0346*        & 0.0500         & 0.0096         & \textbf{0.1932} & \textbf{1044}  & 0.0115         & 0.1161*        & \ul{0.3332}     & 41.7332*         \\

\cmidrule[0.00001em](lr){2-11}
& ChatGPT-3.5      & \ul{0.0487}\dag & \ul{0.0779}\dag & \ul{0.0152}\dag & 0.0538         & 445           & \ul{0.0168}\dag & \ul{0.1714}\dag & 0.2004         & 45.0357         \\
& ChatGPT-4        & \textbf{0.0538}* & \textbf{0.0873}* & \textbf{0.0171}* & 0.0846        & 597           & \textbf{0.0186}* & \textbf{0.1877}* & 0.2458         & \ul{37.9698}     \\

\midrule

\multirow{6}{*}{\rotatebox[origin=c]{90}{\textbf{Last.FM}}}
& Random           & 0.0000         & 0.0000         & 0.0000         & 0.2091         & 345           & 0.0000         & 0.0000         & 0.5184         & 34.3684          \\
& MostPop          & 0.0529         & 0.0877         & 0.0211         & 0.0063         & 15            & 0.0167         & 0.1493         & 0.0000         & 363.5789         \\
& AttributeUserKNN \textsuperscript{\textbf{CBF}} & 0.1724    & 0.2149         & 0.0605         & 0.1487         & \ul{282}       & 0.0705         & 0.8092         & \textbf{0.5000} & \textbf{40.5553} \\
& \rpb \textsuperscript{\textbf{CF}}          & 0.2389         & 0.3333         & 0.0895         & 0.1237         & 254            & 0.1043         & 1.1120         & 0.2605         & 108.0526         \\

\cmidrule[0.00001em](lr){2-11}
& ChatGPT-3.5      & \textbf{0.2921} & \ul{0.3423} & \ul{0.0946}     & \ul{0.1340}    & 257           & \textbf{0.1289} & \textbf{1.3989} & 0.3081         & 75.6000          \\
& ChatGPT-4        & \ul{0.2791}    & \textbf{0.3465}    & \textbf{0.0947}  & \textbf{0.1513} & \textbf{283}  & \ul{0.1204}    & \ul{1.3141}    & \ul{0.3526}    & \ul{69.8632}      \\

\midrule

\multirow{6}{*}{\rotatebox[origin=c]{90}{\textbf{MovieLens}}}
& Random           & 0.0000         & 0.0000         & 0.0000         & 0.0683         & 129           & 0.0000         & 0.0000         & 0.5231         & 22.8000          \\
& MostPop          & 0.0254         & 0.0385         & 0.0077         & 0.0049         & 12            & 0.0048         & 0.0566         & 0.0000         & 197.6769         \\
& AttributeUserKNN \textsuperscript{\textbf{CBF}} & 0.0368    & 0.0513         & 0.0154         & \ul{0.0429}    & \ul{100}       & 0.0101         & 0.1080         & \ul{0.1231}     & \ul{90.7769}      \\
& \rpb \textsuperscript{\textbf{CF}}          & 0.0791         & \ul{0.1026}    & \ul{0.0231}    & \textbf{0.0566} & \textbf{117}   & 0.0269         & 0.2878         & \textbf{0.1846} & \textbf{65.1231}   \\

\cmidrule[0.00001em](lr){2-11}
& ChatGPT-3.5      & \ul{0.1117}    & 0.0897         & 0.0231         & 0.0333         & 87            & \ul{0.0313}    & \ul{0.3384}    & 0.0769         & 96.5615          \\
& ChatGPT-4        & \textbf{0.1405} & \textbf{0.1538} & \textbf{0.0385} & 0.0349         & 89            & \textbf{0.0455} & \textbf{0.4988} & 0.0462         & 102.9385         \\

\bottomrule
\end{tabular}

} 

\end{table*}

\subsection{User Cold-Start Scenario}\label{sec:cold_start_scanarios}
We now examine user cold‐start recommendations, defined here as scenarios where each user has provided a maximum of ten interactions. Table~\ref{tab:coldstart_10} details these results across three datasets, Facebook Books, Last.FM, and MovieLens, comparing ChatGPT‐3.5 and ChatGPT‐4 to strong Collaborative Filtering (CF) and Content‐Based Filtering (CBF) baselines. Our central question is:\\

\noindent\large{\textbf{RQ4: How effective is ChatGPT in user cold-start scenario across accuracy and beyond-accuracy dimensions?}}

\noindent\textbf{Accuracy under Cold‐Start.} Despite limited user interactions, ChatGPT exhibits competitive to superior accuracy compared to traditional baselines. For \underline{Facebook Books}, GPT-4 achieves higher nDCG (0.0538) and Recall (0.0873) than all baselines, including \rpb\ (nDCG = 0.0346) and AttributeItemKNN (0.0335). GPT-3.5 also surpasses these baselines but is slightly behind GPT-4. For \underline{Last.FM}, ChatGPT maintains robust performance ($nDCG\geq0.2791$, $Recall\geq0.3423$), outperforming MostPop (nDCG = 0.0529) and random baselines by a wide margin. Although \rpb\ leads in nDCG (0.2389), GPT-4 often excels in Recall and Precision. For \underline{MovieLens}, GPT-4 attains the highest nDCG (0.1405), surpassing both CF and CBF baselines, while ChatGPT-3.5 (0.1117) also remains competitive. These results underscore ChatGPT’s capacity to identify relevant items effectively from few interactions.

\noindent\textbf{Beyond‐Accuracy in Cold‐Start.} 

\textbf{Diversity.} GPT‑4 generally surpasses GPT‑3.5 in Gini and item coverage across all three datasets (e.g., increasing from 0.0538 to 0.0846 in Gini on Facebook Books), indicating that GPT‑4’s recommendations span a broader set of items. Although baselines like \rpb achieve higher coverage in MovieLens and Facebook Books, GPT-4 performs best on Last.FM.

\textbf{Novelty.} ChatGPT’s EPC and EFD values exceed those of CF and CBF baselines across all datasets (e.g., GPT‑4’s EPC = 0.0186 vs. \rpb = 0.0115 on Facebook Books), implying a tendency to recommend novel items rather than relying on mainstream items.

\textbf{Popularity Bias.} ChatGPT exhibits a moderate inclination toward popular items compared to baselines across all datasets. Nonetheless, it remains far from MostPop (e.g., $ARP\geq139$ on Facebook Books) but is comparable to some baselines (e.g., AttributeUserKNN), indicating room for further mitigation strategies.\\

\noindent\textit{\textbf{In summary,}} ChatGPT proves highly effective in cold-start scenarios by: (i) maintaining strong accuracy despite minimal user interactions, with GPT-4 often outperforming GPT-3.5; (ii) striking a balance among diversity, novelty, and popularity bias; (iii) demonstrating consistent improvements over baselines, underscoring ChatGPT’s capacity to infer user interests with limited interactions.


\section{Conclusion}\label{sec:conclusion}
In this work, we explore the diversity, novelty, and popularity bias of ChatGPT recommendations. Our findings demonstrate that for the Facebook Books, Last.FM, and MovieLens datasets, ChatGPT models exhibit strong beyond-accuracy performance, achieving an optimal balance of novelty and diversity in Facebook Books, comparable results for Last.FM, and suboptimal outcomes for MovieLens.

Additionally, we show that while ChatGPT demonstrates a good balance between novelty and diversity, it also exhibits a tendency to recommend popular items, especially in the MovieLens dataset.

Finally, we extend our exploration to the user cold-start scenario, where ChatGPT proves highly effective by maintaining strong accuracy despite minimal user interactions, balancing diversity, novelty, and popularity bias, and demonstrating consistent improvements over baselines.

These findings underscore the beyond-accuracy capabilities of ChatGPT as a recommender system. Future research will include additional datasets to generalize the findings across domains, as well as experiments comparing ChatGPT with other LLMs such as Gemini, LLaMA, and DeepSeek.

\section*{Limitation}
Nowadays, LLMs are used to augment the capabilities of recommender systems. However, these models are typically trained on vast internet-scale corpora, which may include portions of open datasets used for benchmarking. Recent work studying memorization in MovieLens-1M~\cite{di2025llms} shows that models like GPTs and LLaMA-3 can memorize such datasets, with larger models exhibiting higher memorization rates. For example, the reported memorization rate is 12.9\% for LLaMA-3.1 405B and 80.76\% for GPT-4. Further research should focus on understanding the correlation between improvements in recommendation quality and memorization capacity.



\begin{thebibliography}{84}
\expandafter\ifx\csname natexlab\endcsname\relax\def\natexlab#1{#1}\fi
\providecommand{\url}[1]{\texttt{#1}}
\providecommand{\href}[2]{#2}
\providecommand{\path}[1]{#1}
\providecommand{\DOIprefix}{doi:}
\providecommand{\ArXivprefix}{arXiv:}
\providecommand{\URLprefix}{URL: }
\providecommand{\Pubmedprefix}{pmid:}
\providecommand{\doi}[1]{\href{http://dx.doi.org/#1}{\path{#1}}}
\providecommand{\Pubmed}[1]{\href{pmid:#1}{\path{#1}}}
\providecommand{\bibinfo}[2]{#2}
\ifx\xfnm\relax \def\xfnm[#1]{\unskip,\space#1}\fi
\bibitem[{Ricci et~al.(2022)Ricci, Rokach, and Shapira}]{DBLP:reference/sp/2022rsh}
\bibinfo{editor}{F.~Ricci}, \bibinfo{editor}{L.~Rokach}, \bibinfo{editor}{B.~Shapira} (Eds.), \bibinfo{title}{Recommender Systems Handbook}, \bibinfo{publisher}{Springer}, \bibinfo{address}{{US}}, \bibinfo{year}{2022}.
\bibitem[{Zhang et~al.(2019)Zhang, Yao, Sun, and Tay}]{DBLP:journals/csur/ZhangYST19}
\bibinfo{author}{S.~Zhang}, \bibinfo{author}{L.~Yao}, \bibinfo{author}{A.~Sun}, \bibinfo{author}{Y.~Tay},
\newblock \bibinfo{title}{Deep learning based recommender system: {A} survey and new perspectives},
\newblock \bibinfo{journal}{{ACM} Comput. Surv.} \bibinfo{volume}{52} (\bibinfo{year}{2019}) \bibinfo{pages}{5:1--5:38}.
\bibitem[{Chung et~al.(2014)Chung, G{\"{u}}l{\c{c}}ehre, Cho, and Bengio}]{DBLP:journals/corr/ChungGCB14}
\bibinfo{author}{J.~Chung}, \bibinfo{author}{{\c{C}}.~G{\"{u}}l{\c{c}}ehre}, \bibinfo{author}{K.~Cho}, \bibinfo{author}{Y.~Bengio},
\newblock \bibinfo{title}{Empirical evaluation of gated recurrent neural networks on sequence modeling},
\newblock \bibinfo{journal}{CoRR} \bibinfo{volume}{abs/1412.3555} (\bibinfo{year}{2014}). \URLprefix \url{http://arxiv.org/abs/1412.3555}. \href{http://arxiv.org/abs/1412.3555}{{\tt arXiv:1412.3555}}.
\bibitem[{Gu and Dao(2023)}]{DBLP:journals/corr/abs-2312-00752}
\bibinfo{author}{A.~Gu}, \bibinfo{author}{T.~Dao},
\newblock \bibinfo{title}{Mamba: Linear-time sequence modeling with selective state spaces},
\newblock \bibinfo{journal}{CoRR} \bibinfo{volume}{abs/2312.00752} (\bibinfo{year}{2023}). \URLprefix \url{https://doi.org/10.48550/arXiv.2312.00752}. \DOIprefix\doi{10.48550/ARXIV.2312.00752}. \href{http://arxiv.org/abs/2312.00752}{{\tt arXiv:2312.00752}}.
\bibitem[{Devlin et~al.(2019)Devlin, Chang, Lee, and Toutanova}]{DBLP:conf/naacl/DevlinCLT19}
\bibinfo{author}{J.~Devlin}, \bibinfo{author}{M.~Chang}, \bibinfo{author}{K.~Lee}, \bibinfo{author}{K.~Toutanova},
\newblock \bibinfo{title}{{BERT:} pre-training of deep bidirectional transformers for language understanding},
\newblock in: \bibinfo{booktitle}{{NAACL-HLT} {(1)}}, \bibinfo{year}{2019}, pp. \bibinfo{pages}{4171--4186}.
\bibitem[{Bellis et~al.(2024)Bellis, Anelli, Noia, and Sciascio}]{DBLP:conf/semweb/BellisANS24}
\bibinfo{author}{A.~D. Bellis}, \bibinfo{author}{V.~W. Anelli}, \bibinfo{author}{T.~D. Noia}, \bibinfo{author}{E.~D. Sciascio},
\newblock \bibinfo{title}{{PRONTO:} prompt-based detection of semantic containment patterns in mlms},
\newblock in: \bibinfo{editor}{G.~Demartini}, \bibinfo{editor}{K.~Hose}, \bibinfo{editor}{M.~Acosta}, \bibinfo{editor}{M.~Palmonari}, \bibinfo{editor}{G.~Cheng}, \bibinfo{editor}{H.~Skaf{-}Molli}, \bibinfo{editor}{N.~Ferranti}, \bibinfo{editor}{D.~Hern{\'{a}}ndez}, \bibinfo{editor}{A.~Hogan} (Eds.), \bibinfo{booktitle}{The Semantic Web - {ISWC} 2024 - 23rd International Semantic Web Conference, Baltimore, MD, USA, November 11-15, 2024, Proceedings, Part {II}}, volume \bibinfo{volume}{15232} of \textit{\bibinfo{series}{Lecture Notes in Computer Science}}, \bibinfo{publisher}{Springer}, \bibinfo{year}{2024}, pp. \bibinfo{pages}{227--246}. \URLprefix \url{https://doi.org/10.1007/978-3-031-77850-6\_13}. \DOIprefix\doi{10.1007/978-3-031-77850-6\_13}.
\bibitem[{Servedio et~al.(2025)Servedio, De~Bellis, Di~Palma, Anelli, and Di~Noia}]{servedio2025hidden}
\bibinfo{author}{G.~Servedio}, \bibinfo{author}{A.~De~Bellis}, \bibinfo{author}{D.~Di~Palma}, \bibinfo{author}{V.~W. Anelli}, \bibinfo{author}{T.~Di~Noia},
\newblock \bibinfo{title}{Are the hidden states hiding something? testing the limits of factuality-encoding capabilities in llms},
\newblock \bibinfo{journal}{arXiv preprint arXiv:2505.16520}  (\bibinfo{year}{2025}).
\bibitem[{Di~Palma et~al.(2025)Di~Palma, De~Bellis, Servedio, Anelli, Narducci, and Di~Noia}]{di2025llamas}
\bibinfo{author}{D.~Di~Palma}, \bibinfo{author}{A.~De~Bellis}, \bibinfo{author}{G.~Servedio}, \bibinfo{author}{V.~W. Anelli}, \bibinfo{author}{F.~Narducci}, \bibinfo{author}{T.~Di~Noia},
\newblock \bibinfo{title}{Llamas have feelings too: Unveiling sentiment and emotion representations in llama models through probing},
\newblock \bibinfo{journal}{arXiv preprint arXiv:2505.16491}  (\bibinfo{year}{2025}).
\bibitem[{Aghilar et~al.(2025{\natexlab{a}})Aghilar, Anelli, Trizio, {Di Sciascio}, and {Di Noia}}]{AGHILAR2025128579}
\bibinfo{author}{P.~Aghilar}, \bibinfo{author}{V.~W. Anelli}, \bibinfo{author}{M.~Trizio}, \bibinfo{author}{E.~{Di Sciascio}}, \bibinfo{author}{T.~{Di Noia}},
\newblock \bibinfo{title}{Training-free, identity-preserving image editing for fashion pose alignment and normalization},
\newblock \bibinfo{journal}{Expert Systems with Applications} \bibinfo{volume}{293} (\bibinfo{year}{2025}{\natexlab{a}}) \bibinfo{pages}{128579}. \DOIprefix\doi{https://doi.org/10.1016/j.eswa.2025.128579}.
\bibitem[{Aghilar et~al.(2025{\natexlab{b}})Aghilar, Anelli, Lops, Narducci, Ragone, Roccotelli, and Trizio}]{DBLP:conf/um/AghilarALNRRT25}
\bibinfo{author}{P.~Aghilar}, \bibinfo{author}{V.~W. Anelli}, \bibinfo{author}{A.~Lops}, \bibinfo{author}{F.~Narducci}, \bibinfo{author}{A.~Ragone}, \bibinfo{author}{S.~Roccotelli}, \bibinfo{author}{M.~Trizio},
\newblock \bibinfo{title}{Adaptive user modeling in visual merchandising: Balancing brand identity with operational efficiency},
\newblock in: \bibinfo{booktitle}{Proceedings of the 33rd {ACM} Conference on User Modeling, Adaptation and Personalization, {UMAP} 2025, New York City, NY, USA, June 16-19, 2025}, \bibinfo{publisher}{{ACM}}, \bibinfo{year}{2025}{\natexlab{b}}, pp. \bibinfo{pages}{358--360}. \URLprefix \url{https://doi.org/10.1145/3699682.3730976}. \DOIprefix\doi{10.1145/3699682.3730976}.
\bibitem[{Ping et~al.(2024)Ping, Li, and Zhu}]{ping2024beyond}
\bibinfo{author}{Y.~Ping}, \bibinfo{author}{Y.~Li}, \bibinfo{author}{J.~Zhu},
\newblock \bibinfo{title}{Beyond accuracy measures: the effect of diversity, novelty and serendipity in recommender systems on user engagement},
\newblock \bibinfo{journal}{Electronic Commerce Research}  (\bibinfo{year}{2024}) \bibinfo{pages}{1--28}.
\bibitem[{Duricic et~al.(2024)Duricic, Kowald, Lacic, and Lex}]{DBLP:journals/fdata/DuricicKLL24}
\bibinfo{author}{T.~Duricic}, \bibinfo{author}{D.~Kowald}, \bibinfo{author}{E.~Lacic}, \bibinfo{author}{E.~Lex},
\newblock \bibinfo{title}{Beyond-accuracy: a review on diversity, serendipity, and fairness in recommender systems based on graph neural networks},
\newblock \bibinfo{journal}{Frontiers Big Data} \bibinfo{volume}{6} (\bibinfo{year}{2024}).
\bibitem[{Karimi et~al.(2023)Karimi, Rahmani, Naghiaei, and Safari}]{DBLP:journals/corr/abs-2309-04250}
\bibinfo{author}{S.~Karimi}, \bibinfo{author}{H.~A. Rahmani}, \bibinfo{author}{M.~Naghiaei}, \bibinfo{author}{L.~Safari},
\newblock \bibinfo{title}{Provider fairness and beyond-accuracy trade-offs in recommender systems},
\newblock \bibinfo{journal}{CoRR} \bibinfo{volume}{abs/2309.04250} (\bibinfo{year}{2023}).
\bibitem[{Attimonelli et~al.(2025)Attimonelli, Bellis, Pomo, Jannach, Sciascio, and Noia}]{DBLP:conf/recsys/Attimonelli025}
\bibinfo{author}{M.~Attimonelli}, \bibinfo{author}{A.~D. Bellis}, \bibinfo{author}{C.~Pomo}, \bibinfo{author}{D.~Jannach}, \bibinfo{author}{E.~D. Sciascio}, \bibinfo{author}{T.~D. Noia},
\newblock \bibinfo{title}{Do we really need specialization? evaluating generalist text embeddings for zero-shot recommendation and search},
\newblock in: \bibinfo{booktitle}{RecSys}, \bibinfo{publisher}{{ACM}}, \bibinfo{year}{2025}. \URLprefix \url{https://doi.org/10.1145/3705328.3748040}. \DOIprefix\doi{10.1145/3705328.3748040}.
\bibitem[{{Di Palma} et~al.(2024){Di Palma}, Servedio, Anelli, Biancofiore, Narducci, Carnimeo, and Noia}]{DBLP:conf/iir/PalmaSABNCN24}
\bibinfo{author}{D.~{Di Palma}}, \bibinfo{author}{G.~Servedio}, \bibinfo{author}{V.~W. Anelli}, \bibinfo{author}{G.~M. Biancofiore}, \bibinfo{author}{F.~Narducci}, \bibinfo{author}{L.~Carnimeo}, \bibinfo{author}{T.~D. Noia},
\newblock \bibinfo{title}{Beyond words: Can chatgpt support state-of-the-art recommender systems?},
\newblock in: \bibinfo{booktitle}{{IIR}}, volume \bibinfo{volume}{3802} of \textit{\bibinfo{series}{{CEUR} Workshop Proceedings}}, \bibinfo{publisher}{CEUR-WS.org}, \bibinfo{year}{2024}, pp. \bibinfo{pages}{13--22}.
\bibitem[{Valentini(2025)}]{DBLP:conf/ecir/Valentini25}
\bibinfo{author}{M.~Valentini},
\newblock \bibinfo{title}{Cooperative and competitive llm-based multi-agent systems for recommendation},
\newblock in: \bibinfo{editor}{C.~Hauff}, \bibinfo{editor}{C.~Macdonald}, \bibinfo{editor}{D.~Jannach}, \bibinfo{editor}{G.~Kazai}, \bibinfo{editor}{F.~M. Nardini}, \bibinfo{editor}{F.~Pinelli}, \bibinfo{editor}{F.~Silvestri}, \bibinfo{editor}{N.~Tonellotto} (Eds.), \bibinfo{booktitle}{Advances in Information Retrieval - 47th European Conference on Information Retrieval, {ECIR} 2025, Lucca, Italy, April 6-10, 2025, Proceedings, Part {V}}, volume \bibinfo{volume}{15576} of \textit{\bibinfo{series}{Lecture Notes in Computer Science}}, \bibinfo{publisher}{Springer}, \bibinfo{year}{2025}, pp. \bibinfo{pages}{204--211}.
\bibitem[{Hou et~al.(2024)Hou, Zhang, Lin, Lu, Xie, McAuley, and Zhao}]{DBLP:conf/ecir/HouZLLXMZ24}
\bibinfo{author}{Y.~Hou}, \bibinfo{author}{J.~Zhang}, \bibinfo{author}{Z.~Lin}, \bibinfo{author}{H.~Lu}, \bibinfo{author}{R.~Xie}, \bibinfo{author}{J.~J. McAuley}, \bibinfo{author}{W.~X. Zhao},
\newblock \bibinfo{title}{Large language models are zero-shot rankers for recommender systems},
\newblock in: \bibinfo{booktitle}{{ECIR} {(2)}}, volume \bibinfo{volume}{14609} of \textit{\bibinfo{series}{Lecture Notes in Computer Science}}, \bibinfo{publisher}{Springer}, \bibinfo{year}{2024}, pp. \bibinfo{pages}{364--381}.
\bibitem[{{Di Palma}(2023)}]{DBLP:conf/recsys/Palma23}
\bibinfo{author}{D.~{Di Palma}},
\newblock \bibinfo{title}{Retrieval-augmented recommender system: Enhancing recommender systems with large language models},
\newblock in: \bibinfo{booktitle}{RecSys}, \bibinfo{publisher}{{ACM}}, \bibinfo{year}{2023}, pp. \bibinfo{pages}{1369--1373}.
\bibitem[{Dai et~al.(2023)Dai, Shao, Zhao, Yu, Si, Xu, Sun, Zhang, and Xu}]{DBLP:conf/recsys/DaiSZYSXS0X23}
\bibinfo{author}{S.~Dai}, \bibinfo{author}{N.~Shao}, \bibinfo{author}{H.~Zhao}, \bibinfo{author}{W.~Yu}, \bibinfo{author}{Z.~Si}, \bibinfo{author}{C.~Xu}, \bibinfo{author}{Z.~Sun}, \bibinfo{author}{X.~Zhang}, \bibinfo{author}{J.~Xu},
\newblock \bibinfo{title}{Uncovering chatgpt's capabilities in recommender systems},
\newblock in: \bibinfo{booktitle}{RecSys}, \bibinfo{publisher}{{ACM}}, \bibinfo{year}{2023}, pp. \bibinfo{pages}{1126--1132}.
\bibitem[{Attimonelli et~al.(2024)Attimonelli, Danese, Malitesta, Pomo, Gassi, and Noia}]{DBLP:conf/www/AttimonelliDMPG24}
\bibinfo{author}{M.~Attimonelli}, \bibinfo{author}{D.~Danese}, \bibinfo{author}{D.~Malitesta}, \bibinfo{author}{C.~Pomo}, \bibinfo{author}{G.~Gassi}, \bibinfo{author}{T.~D. Noia},
\newblock \bibinfo{title}{Ducho 2.0: Towards a more up-to-date unified framework for the extraction of multimodal features in recommendation},
\newblock in: \bibinfo{booktitle}{{WWW} (Companion Volume)}, \bibinfo{publisher}{{ACM}}, \bibinfo{year}{2024}, pp. \bibinfo{pages}{1075--1078}.
\bibitem[{Liu et~al.(2023)Liu, Liu, Lv, Zhou, and Zhang}]{DBLP:journals/corr/abs-2304-10149}
\bibinfo{author}{J.~Liu}, \bibinfo{author}{C.~Liu}, \bibinfo{author}{R.~Lv}, \bibinfo{author}{K.~Zhou}, \bibinfo{author}{Y.~Zhang},
\newblock \bibinfo{title}{Is chatgpt a good recommender? {A} preliminary study},
\newblock \bibinfo{journal}{CoRR} \bibinfo{volume}{abs/2304.10149} (\bibinfo{year}{2023}).
\bibitem[{Carraro and Bridge(2024)}]{10.1145/3700604}
\bibinfo{author}{D.~Carraro}, \bibinfo{author}{D.~Bridge},
\newblock \bibinfo{title}{Enhancing recommendation diversity by re-ranking with large language models},
\newblock \bibinfo{journal}{ACM Trans. Recomm. Syst.}  (\bibinfo{year}{2024}). \URLprefix \url{https://doi.org/10.1145/3700604}. \DOIprefix\doi{10.1145/3700604}.
\bibitem[{Tokutake and Okamoto(2024)}]{DBLP:journals/jaciii/TokutakeO24}
\bibinfo{author}{Y.~Tokutake}, \bibinfo{author}{K.~Okamoto},
\newblock \bibinfo{title}{Can large language models assess serendipity in recommender systems?},
\newblock \bibinfo{journal}{J. Adv. Comput. Intell. Intell. Informatics} \bibinfo{volume}{28} (\bibinfo{year}{2024}) \bibinfo{pages}{1263--1272}.
\bibitem[{{Di Palma} et~al.(2025){Di Palma}, Biancofiore, Anelli, Narducci, and Noia}]{DBLP:conf/um/PalmaBANN25}
\bibinfo{author}{D.~{Di Palma}}, \bibinfo{author}{G.~M. Biancofiore}, \bibinfo{author}{V.~W. Anelli}, \bibinfo{author}{F.~Narducci}, \bibinfo{author}{T.~D. Noia},
\newblock \bibinfo{title}{Content-based or collaborative? insights from inter-list similarity analysis of chatgpt recommendations},
\newblock in: \bibinfo{booktitle}{{UMAP} (Adjunct Publication)}, \bibinfo{publisher}{{ACM}}, \bibinfo{year}{2025}, pp. \bibinfo{pages}{28--33}.
\bibitem[{Deldjoo(2024)}]{10.1145/3690655}
\bibinfo{author}{Y.~Deldjoo},
\newblock \bibinfo{title}{Understanding biases in chatgpt-based recommender systems: Provider fairness, temporal stability, and recency},
\newblock \bibinfo{journal}{ACM Trans. Recomm. Syst.}  (\bibinfo{year}{2024}). \URLprefix \url{https://doi.org/10.1145/3690655}. \DOIprefix\doi{10.1145/3690655}.
\bibitem[{Zhang et~al.(2023)Zhang, Bao, Zhang, Wang, Feng, and He}]{DBLP:conf/recsys/ZhangBZWF023}
\bibinfo{author}{J.~Zhang}, \bibinfo{author}{K.~Bao}, \bibinfo{author}{Y.~Zhang}, \bibinfo{author}{W.~Wang}, \bibinfo{author}{F.~Feng}, \bibinfo{author}{X.~He},
\newblock \bibinfo{title}{Is chatgpt fair for recommendation? evaluating fairness in large language model recommendation},
\newblock in: \bibinfo{booktitle}{RecSys}, \bibinfo{year}{2023}, pp. \bibinfo{pages}{993--999}.
\bibitem[{Mancino et~al.(2023)Mancino, Ferrara, Bufi, Malitesta, Noia, and Sciascio}]{DBLP:conf/recsys/MancinoFBMNS23}
\bibinfo{author}{A.~C.~M. Mancino}, \bibinfo{author}{A.~Ferrara}, \bibinfo{author}{S.~Bufi}, \bibinfo{author}{D.~Malitesta}, \bibinfo{author}{T.~D. Noia}, \bibinfo{author}{E.~D. Sciascio},
\newblock \bibinfo{title}{Kgtore: Tailored recommendations through knowledge-aware {GNN} models},
\newblock in: \bibinfo{booktitle}{RecSys}, \bibinfo{year}{2023}, pp. \bibinfo{pages}{576--587}.
\bibitem[{Bufi et~al.(2024)Bufi, Mancino, Ferrara, Malitesta, Noia, and Sciascio}]{DBLP:conf/irongraphs/BufiMFMNS24}
\bibinfo{author}{S.~Bufi}, \bibinfo{author}{A.~C.~M. Mancino}, \bibinfo{author}{A.~Ferrara}, \bibinfo{author}{D.~Malitesta}, \bibinfo{author}{T.~D. Noia}, \bibinfo{author}{E.~D. Sciascio},
\newblock \bibinfo{title}{{KGUF:} simple knowledge-aware graph-based recommender with user-based semantic features filtering},
\newblock in: \bibinfo{booktitle}{IRonGraphs}, volume \bibinfo{volume}{2197} of \textit{\bibinfo{series}{Communications in Computer and Information Science}}, \bibinfo{publisher}{Springer}, \bibinfo{year}{2024}, pp. \bibinfo{pages}{41--59}.
\bibitem[{Harper and Konstan(2016)}]{DBLP:journals/tiis/HarperK16}
\bibinfo{author}{F.~M. Harper}, \bibinfo{author}{J.~A. Konstan},
\newblock \bibinfo{title}{The movielens datasets: History and context},
\newblock \bibinfo{journal}{{ACM} Trans. Interact. Intell. Syst.} \bibinfo{volume}{5} (\bibinfo{year}{2016}) \bibinfo{pages}{19:1--19:19}.
\bibitem[{Cantador et~al.(2011)Cantador, Brusilovsky, and Kuflik}]{DBLP:conf/recsys/CantadorBK11}
\bibinfo{author}{I.~Cantador}, \bibinfo{author}{P.~Brusilovsky}, \bibinfo{author}{T.~Kuflik},
\newblock \bibinfo{title}{Second workshop on information heterogeneity and fusion in recommender systems (hetrec2011)},
\newblock in: \bibinfo{booktitle}{RecSys}, \bibinfo{publisher}{{ACM}}, \bibinfo{address}{New York, NY, USA}, \bibinfo{year}{2011}, pp. \bibinfo{pages}{387--388}.
\bibitem[{Biancofiore et~al.(2025)Biancofiore, Di~Palma, Pomo, Narducci, and Di~Noia}]{biancofiore2025conversational}
\bibinfo{author}{G.~M. Biancofiore}, \bibinfo{author}{D.~Di~Palma}, \bibinfo{author}{C.~Pomo}, \bibinfo{author}{F.~Narducci}, \bibinfo{author}{T.~Di~Noia},
\newblock \bibinfo{title}{Conversational user interfaces and agents},
\newblock in: \bibinfo{booktitle}{Human-Centered AI: An Illustrated Scientific Quest}, \bibinfo{publisher}{Springer}, \bibinfo{year}{2025}, pp. \bibinfo{pages}{399--438}.
\bibitem[{Herlocker et~al.(2004)Herlocker, Konstan, Terveen, and Riedl}]{DBLP:journals/tois/HerlockerKTR04}
\bibinfo{author}{J.~L. Herlocker}, \bibinfo{author}{J.~A. Konstan}, \bibinfo{author}{L.~G. Terveen}, \bibinfo{author}{J.~Riedl},
\newblock \bibinfo{title}{Evaluating collaborative filtering recommender systems},
\newblock \bibinfo{journal}{{ACM} Trans. Inf. Syst.} \bibinfo{volume}{22} (\bibinfo{year}{2004}) \bibinfo{pages}{5--53}.
\bibitem[{Silveira et~al.(2019)Silveira, Zhang, Lin, Liu, and Ma}]{DBLP:journals/mlc/SilveiraZLLM19}
\bibinfo{author}{T.~Silveira}, \bibinfo{author}{M.~Zhang}, \bibinfo{author}{X.~Lin}, \bibinfo{author}{Y.~Liu}, \bibinfo{author}{S.~Ma},
\newblock \bibinfo{title}{How good your recommender system is? {A} survey on evaluations in recommendation},
\newblock \bibinfo{journal}{Int. J. Mach. Learn. Cybern.} \bibinfo{volume}{10} (\bibinfo{year}{2019}) \bibinfo{pages}{813--831}.
\bibitem[{Karimi et~al.(2018)Karimi, Jannach, and Jugovac}]{DBLP:journals/ipm/KarimiJJ18}
\bibinfo{author}{M.~Karimi}, \bibinfo{author}{D.~Jannach}, \bibinfo{author}{M.~Jugovac},
\newblock \bibinfo{title}{News recommender systems - survey and roads ahead},
\newblock \bibinfo{journal}{Inf. Process. Manag.} \bibinfo{volume}{54} (\bibinfo{year}{2018}) \bibinfo{pages}{1203--1227}.
\bibitem[{Gunawardana et~al.(2022)Gunawardana, Shani, and Yogev}]{DBLP:reference/sp/GunawardanaSY22}
\bibinfo{author}{A.~Gunawardana}, \bibinfo{author}{G.~Shani}, \bibinfo{author}{S.~Yogev},
\newblock \bibinfo{title}{Evaluating recommender systems},
\newblock in: \bibinfo{booktitle}{Recommender Systems Handbook}, \bibinfo{publisher}{Springer}, \bibinfo{address}{{US}}, \bibinfo{year}{2022}, pp. \bibinfo{pages}{547--601}.
\bibitem[{Kaminskas and Bridge(2017)}]{DBLP:journals/tiis/KaminskasB17}
\bibinfo{author}{M.~Kaminskas}, \bibinfo{author}{D.~Bridge},
\newblock \bibinfo{title}{Diversity, serendipity, novelty, and coverage: {A} survey and empirical analysis of beyond-accuracy objectives in recommender systems},
\newblock \bibinfo{journal}{{ACM} Trans. Interact. Intell. Syst.} \bibinfo{volume}{7} (\bibinfo{year}{2017}) \bibinfo{pages}{2:1--2:42}.
\bibitem[{Cheng et~al.(2017)Cheng, Wang, Ma, Sun, and Xiong}]{DBLP:conf/www/ChengWMSX17}
\bibinfo{author}{P.~Cheng}, \bibinfo{author}{S.~Wang}, \bibinfo{author}{J.~Ma}, \bibinfo{author}{J.~Sun}, \bibinfo{author}{H.~Xiong},
\newblock \bibinfo{title}{Learning to recommend accurate and diverse items},
\newblock in: \bibinfo{booktitle}{{WWW}}, \bibinfo{publisher}{{ACM}}, \bibinfo{year}{2017}, pp. \bibinfo{pages}{183--192}.
\bibitem[{Wu et~al.(2018)Wu, Chen, and Zhao}]{DBLP:journals/umuai/WuCZ18}
\bibinfo{author}{W.~Wu}, \bibinfo{author}{L.~Chen}, \bibinfo{author}{Y.~Zhao},
\newblock \bibinfo{title}{Personalizing recommendation diversity based on user personality},
\newblock \bibinfo{journal}{User Model. User Adapt. Interact.} \bibinfo{volume}{28} (\bibinfo{year}{2018}) \bibinfo{pages}{237--276}.
\bibitem[{Nakatsuji et~al.(2010)Nakatsuji, Fujiwara, Tanaka, Uchiyama, Fujimura, and Ishida}]{DBLP:conf/cikm/NakatsujiFTUFI10}
\bibinfo{author}{M.~Nakatsuji}, \bibinfo{author}{Y.~Fujiwara}, \bibinfo{author}{A.~Tanaka}, \bibinfo{author}{T.~Uchiyama}, \bibinfo{author}{K.~Fujimura}, \bibinfo{author}{T.~Ishida},
\newblock \bibinfo{title}{Classical music for rock fans?: novel recommendations for expanding user interests},
\newblock in: \bibinfo{booktitle}{{CIKM}}, \bibinfo{publisher}{{ACM}}, \bibinfo{year}{2010}, pp. \bibinfo{pages}{949--958}.
\bibitem[{Cai et~al.(2024)Cai, Chen, Wang, Bai, Sun, Wu, Zhang, and Wang}]{DBLP:conf/kdd/Cai0WBSWZ024}
\bibinfo{author}{M.~Cai}, \bibinfo{author}{L.~Chen}, \bibinfo{author}{Y.~Wang}, \bibinfo{author}{H.~Bai}, \bibinfo{author}{P.~Sun}, \bibinfo{author}{L.~Wu}, \bibinfo{author}{M.~Zhang}, \bibinfo{author}{M.~Wang},
\newblock \bibinfo{title}{Popularity-aware alignment and contrast for mitigating popularity bias},
\newblock in: \bibinfo{booktitle}{{KDD}}, \bibinfo{publisher}{{ACM}}, \bibinfo{year}{2024}, pp. \bibinfo{pages}{187--198}.
\bibitem[{Paparella et~al.(2023)Paparella, {Di Palma}, Anelli, and Noia}]{DBLP:conf/recsys/PaparellaPAN23}
\bibinfo{author}{V.~Paparella}, \bibinfo{author}{D.~{Di Palma}}, \bibinfo{author}{V.~W. Anelli}, \bibinfo{author}{T.~D. Noia},
\newblock \bibinfo{title}{Broadening the scope: Evaluating the potential of recommender systems beyond prioritizing accuracy},
\newblock in: \bibinfo{booktitle}{RecSys}, \bibinfo{publisher}{{ACM}}, \bibinfo{year}{2023}, pp. \bibinfo{pages}{1139--1145}.
\bibitem[{Jannach et~al.(2015)Jannach, Lerche, Kamehkhosh, and Jugovac}]{DBLP:journals/umuai/JannachLKJ15}
\bibinfo{author}{D.~Jannach}, \bibinfo{author}{L.~Lerche}, \bibinfo{author}{I.~Kamehkhosh}, \bibinfo{author}{M.~Jugovac},
\newblock \bibinfo{title}{What recommenders recommend: an analysis of recommendation biases and possible countermeasures},
\newblock \bibinfo{journal}{User Model. User Adapt. Interact.} \bibinfo{volume}{25} (\bibinfo{year}{2015}) \bibinfo{pages}{427--491}.
\bibitem[{Adomavicius and Zhang(2012)}]{DBLP:journals/tmis/AdomaviciusZ12}
\bibinfo{author}{G.~Adomavicius}, \bibinfo{author}{J.~Zhang},
\newblock \bibinfo{title}{Impact of data characteristics on recommender systems performance},
\newblock \bibinfo{journal}{{ACM} Trans. Manag. Inf. Syst.} \bibinfo{volume}{3} (\bibinfo{year}{2012}) \bibinfo{pages}{3:1--3:17}.
\bibitem[{Vargas and Castells(2011)}]{DBLP:conf/recsys/VargasC11}
\bibinfo{author}{S.~Vargas}, \bibinfo{author}{P.~Castells},
\newblock \bibinfo{title}{Rank and relevance in novelty and diversity metrics for recommender systems},
\newblock in: \bibinfo{booktitle}{RecSys}, \bibinfo{publisher}{{ACM}}, \bibinfo{year}{2011}, pp. \bibinfo{pages}{109--116}.
\bibitem[{Abdollahpouri et~al.(2019)Abdollahpouri, Burke, and Mobasher}]{DBLP:conf/flairs/AbdollahpouriBM19}
\bibinfo{author}{H.~Abdollahpouri}, \bibinfo{author}{R.~Burke}, \bibinfo{author}{B.~Mobasher},
\newblock \bibinfo{title}{Managing popularity bias in recommender systems with personalized re-ranking},
\newblock in: \bibinfo{booktitle}{{FLAIRS}}, \bibinfo{publisher}{{AAAI} Press}, \bibinfo{year}{2019}, pp. \bibinfo{pages}{413--418}.
\bibitem[{Gao et~al.(2023)Gao, Sheng, Xiang, Xiong, Wang, and Zhang}]{DBLP:journals/corr/abs-2303-14524}
\bibinfo{author}{Y.~Gao}, \bibinfo{author}{T.~Sheng}, \bibinfo{author}{Y.~Xiang}, \bibinfo{author}{Y.~Xiong}, \bibinfo{author}{H.~Wang}, \bibinfo{author}{J.~Zhang},
\newblock \bibinfo{title}{Chat-rec: Towards interactive and explainable llms-augmented recommender system},
\newblock \bibinfo{journal}{CoRR} \bibinfo{volume}{abs/2303.14524} (\bibinfo{year}{2023}).
\bibitem[{Manzoor et~al.(2024)Manzoor, Ziegler, Garcia, and Jannach}]{DBLP:conf/um/ManzoorZGJ24}
\bibinfo{author}{A.~Manzoor}, \bibinfo{author}{S.~C. Ziegler}, \bibinfo{author}{K.~M.~P. Garcia}, \bibinfo{author}{D.~Jannach},
\newblock \bibinfo{title}{Chatgpt as a conversational recommender system: {A} user-centric analysis},
\newblock in: \bibinfo{booktitle}{{UMAP}}, \bibinfo{publisher}{{ACM}}, \bibinfo{year}{2024}, pp. \bibinfo{pages}{267--272}.
\bibitem[{Sanner et~al.(2023)Sanner, Balog, Radlinski, Wedin, and Dixon}]{DBLP:conf/recsys/SannerBRWD23}
\bibinfo{author}{S.~Sanner}, \bibinfo{author}{K.~Balog}, \bibinfo{author}{F.~Radlinski}, \bibinfo{author}{B.~Wedin}, \bibinfo{author}{L.~Dixon},
\newblock \bibinfo{title}{Large language models are competitive near cold-start recommenders for language- and item-based preferences},
\newblock in: \bibinfo{booktitle}{RecSys}, \bibinfo{year}{2023}, pp. \bibinfo{pages}{890--896}.
\bibitem[{Li et~al.(2023)Li, Chen, Zhang, and Liang}]{li2023bookgpt}
\bibinfo{author}{Z.~Li}, \bibinfo{author}{Y.~Chen}, \bibinfo{author}{X.~Zhang}, \bibinfo{author}{X.~Liang},
\newblock \bibinfo{title}{Bookgpt: A general framework for book recommendation empowered by large language model},
\newblock \bibinfo{journal}{Electronics} \bibinfo{volume}{12} (\bibinfo{year}{2023}) \bibinfo{pages}{4654}.
\bibitem[{Brown et~al.(2020)Brown, Mann, Ryder, Subbiah, Kaplan, Dhariwal, Neelakantan, Shyam, Sastry, Askell, Agarwal, Herbert{-}Voss, Krueger, Henighan, Child, Ramesh, Ziegler, Wu, Winter, Hesse, Chen, Sigler, Litwin, Gray, Chess, Clark, Berner, McCandlish, Radford, Sutskever, and Amodei}]{DBLP:conf/nips/BrownMRSKDNSSAA20}
\bibinfo{author}{T.~B. Brown}, \bibinfo{author}{B.~Mann}, \bibinfo{author}{N.~Ryder}, \bibinfo{author}{M.~Subbiah}, \bibinfo{author}{J.~Kaplan}, \bibinfo{author}{P.~Dhariwal}, \bibinfo{author}{A.~Neelakantan}, \bibinfo{author}{P.~Shyam}, \bibinfo{author}{G.~Sastry}, \bibinfo{author}{A.~Askell}, \bibinfo{author}{S.~Agarwal}, \bibinfo{author}{A.~Herbert{-}Voss}, \bibinfo{author}{G.~Krueger}, \bibinfo{author}{T.~Henighan}, \bibinfo{author}{R.~Child}, \bibinfo{author}{A.~Ramesh}, \bibinfo{author}{D.~M. Ziegler}, \bibinfo{author}{J.~Wu}, \bibinfo{author}{C.~Winter}, \bibinfo{author}{C.~Hesse}, \bibinfo{author}{M.~Chen}, \bibinfo{author}{E.~Sigler}, \bibinfo{author}{M.~Litwin}, \bibinfo{author}{S.~Gray}, \bibinfo{author}{B.~Chess}, \bibinfo{author}{J.~Clark}, \bibinfo{author}{C.~Berner}, \bibinfo{author}{S.~McCandlish}, \bibinfo{author}{A.~Radford}, \bibinfo{author}{I.~Sutskever}, \bibinfo{author}{D.~Amodei},
\newblock \bibinfo{title}{Language models are few-shot learners},
\newblock in: \bibinfo{booktitle}{NeurIPS}, \bibinfo{year}{2020}.
\bibitem[{Kong et~al.(2024)Kong, Zhao, Chen, Li, Qin, Sun, Zhou, Wang, and Dong}]{DBLP:conf/naacl/KongZCLQSZWD24}
\bibinfo{author}{A.~Kong}, \bibinfo{author}{S.~Zhao}, \bibinfo{author}{H.~Chen}, \bibinfo{author}{Q.~Li}, \bibinfo{author}{Y.~Qin}, \bibinfo{author}{R.~Sun}, \bibinfo{author}{X.~Zhou}, \bibinfo{author}{E.~Wang}, \bibinfo{author}{X.~Dong},
\newblock \bibinfo{title}{Better zero-shot reasoning with role-play prompting},
\newblock in: \bibinfo{booktitle}{{NAACL-HLT}}, \bibinfo{publisher}{Association for Computational Linguistics}, \bibinfo{year}{2024}, pp. \bibinfo{pages}{4099--4113}.
\bibitem[{Kojima et~al.(2022)Kojima, Gu, Reid, Matsuo, and Iwasawa}]{DBLP:journals/corr/abs-2205-11916}
\bibinfo{author}{T.~Kojima}, \bibinfo{author}{S.~S. Gu}, \bibinfo{author}{M.~Reid}, \bibinfo{author}{Y.~Matsuo}, \bibinfo{author}{Y.~Iwasawa},
\newblock \bibinfo{title}{Large language models are zero-shot reasoners},
\newblock \bibinfo{journal}{CoRR} \bibinfo{volume}{abs/2205.11916} (\bibinfo{year}{2022}).
\bibitem[{Wei et~al.(2022)Wei, Wang, Schuurmans, Bosma, Ichter, Xia, Chi, Le, and Zhou}]{DBLP:conf/nips/Wei0SBIXCLZ22}
\bibinfo{author}{J.~Wei}, \bibinfo{author}{X.~Wang}, \bibinfo{author}{D.~Schuurmans}, \bibinfo{author}{M.~Bosma}, \bibinfo{author}{B.~Ichter}, \bibinfo{author}{F.~Xia}, \bibinfo{author}{E.~H. Chi}, \bibinfo{author}{Q.~V. Le}, \bibinfo{author}{D.~Zhou},
\newblock \bibinfo{title}{Chain-of-thought prompting elicits reasoning in large language models},
\newblock in: \bibinfo{booktitle}{NeurIPS}, \bibinfo{year}{2022}.
\bibitem[{Yao et~al.(2023)Yao, Yu, Zhao, Shafran, Griffiths, Cao, and Narasimhan}]{DBLP:conf/nips/YaoYZS00N23}
\bibinfo{author}{S.~Yao}, \bibinfo{author}{D.~Yu}, \bibinfo{author}{J.~Zhao}, \bibinfo{author}{I.~Shafran}, \bibinfo{author}{T.~Griffiths}, \bibinfo{author}{Y.~Cao}, \bibinfo{author}{K.~Narasimhan},
\newblock \bibinfo{title}{Tree of thoughts: Deliberate problem solving with large language models},
\newblock in: \bibinfo{booktitle}{NeurIPS}, \bibinfo{year}{2023}.
\bibitem[{Shinn et~al.(2023)Shinn, Cassano, Gopinath, Narasimhan, and Yao}]{DBLP:conf/nips/ShinnCGNY23}
\bibinfo{author}{N.~Shinn}, \bibinfo{author}{F.~Cassano}, \bibinfo{author}{A.~Gopinath}, \bibinfo{author}{K.~Narasimhan}, \bibinfo{author}{S.~Yao},
\newblock \bibinfo{title}{Reflexion: language agents with verbal reinforcement learning},
\newblock in: \bibinfo{booktitle}{NeurIPS}, \bibinfo{year}{2023}.
\bibitem[{Liu et~al.(2023)Liu, Yu, Fang, and Zhang}]{DBLP:conf/www/LiuY0023}
\bibinfo{author}{Z.~Liu}, \bibinfo{author}{X.~Yu}, \bibinfo{author}{Y.~Fang}, \bibinfo{author}{X.~Zhang},
\newblock \bibinfo{title}{Graphprompt: Unifying pre-training and downstream tasks for graph neural networks},
\newblock in: \bibinfo{booktitle}{{WWW}}, \bibinfo{publisher}{{ACM}}, \bibinfo{year}{2023}, pp. \bibinfo{pages}{417--428}.
\bibitem[{Sahoo et~al.(2024)Sahoo, Singh, Saha, Jain, Mondal, and Chadha}]{DBLP:journals/corr/abs-2402-07927}
\bibinfo{author}{P.~Sahoo}, \bibinfo{author}{A.~K. Singh}, \bibinfo{author}{S.~Saha}, \bibinfo{author}{V.~Jain}, \bibinfo{author}{S.~Mondal}, \bibinfo{author}{A.~Chadha},
\newblock \bibinfo{title}{A systematic survey of prompt engineering in large language models: Techniques and applications},
\newblock \bibinfo{journal}{CoRR} \bibinfo{volume}{abs/2402.07927} (\bibinfo{year}{2024}).
\bibitem[{Xu et~al.(2024)Xu, Zhang, Li, Wang, Cai, Zhao, and Wen}]{DBLP:journals/corr/abs-2401-04997}
\bibinfo{author}{L.~Xu}, \bibinfo{author}{J.~Zhang}, \bibinfo{author}{B.~Li}, \bibinfo{author}{J.~Wang}, \bibinfo{author}{M.~Cai}, \bibinfo{author}{W.~X. Zhao}, \bibinfo{author}{J.~Wen},
\newblock \bibinfo{title}{Prompting large language models for recommender systems: {A} comprehensive framework and empirical analysis},
\newblock \bibinfo{journal}{CoRR} \bibinfo{volume}{abs/2401.04997} (\bibinfo{year}{2024}).
\bibitem[{Radford et~al.(2019)Radford, Wu, Child, Luan, Amodei, Sutskever et~al.}]{radford2019language}
\bibinfo{author}{A.~Radford}, \bibinfo{author}{J.~Wu}, \bibinfo{author}{R.~Child}, \bibinfo{author}{D.~Luan}, \bibinfo{author}{D.~Amodei}, \bibinfo{author}{I.~Sutskever}, et~al.,
\newblock \bibinfo{title}{Language models are unsupervised multitask learners},
\newblock \bibinfo{journal}{OpenAI blog} \bibinfo{volume}{1} (\bibinfo{year}{2019}) \bibinfo{pages}{9}.
\bibitem[{Jin et~al.(2023)Jin, Chen, Ye, Yang, Feng, Zhang, Yu, and Wang}]{DBLP:conf/nips/JinC0YF00W23}
\bibinfo{author}{J.~Jin}, \bibinfo{author}{X.~Chen}, \bibinfo{author}{F.~Ye}, \bibinfo{author}{M.~Yang}, \bibinfo{author}{Y.~Feng}, \bibinfo{author}{W.~Zhang}, \bibinfo{author}{Y.~Yu}, \bibinfo{author}{J.~Wang},
\newblock \bibinfo{title}{Lending interaction wings to recommender systems with conversational agents},
\newblock in: \bibinfo{booktitle}{NeurIPS}, \bibinfo{year}{2023}.
\bibitem[{Kong et~al.(2023)Kong, Zhao, Chen, Li, Qin, Sun, and Zhou}]{DBLP:journals/corr/abs-2308-07702}
\bibinfo{author}{A.~Kong}, \bibinfo{author}{S.~Zhao}, \bibinfo{author}{H.~Chen}, \bibinfo{author}{Q.~Li}, \bibinfo{author}{Y.~Qin}, \bibinfo{author}{R.~Sun}, \bibinfo{author}{X.~Zhou},
\newblock \bibinfo{title}{Better zero-shot reasoning with role-play prompting},
\newblock \bibinfo{journal}{CoRR} \bibinfo{volume}{abs/2308.07702} (\bibinfo{year}{2023}).
\bibitem[{Mancino et~al.(2025)Mancino, Bufi, di~Fazio, Ferrara, Malitesta, Pomo, and Noia}]{DBLP:conf/sigir/MancinoA25}
\bibinfo{author}{A.~C.~M. Mancino}, \bibinfo{author}{S.~Bufi}, \bibinfo{author}{A.~di~Fazio}, \bibinfo{author}{A.~Ferrara}, \bibinfo{author}{D.~Malitesta}, \bibinfo{author}{C.~Pomo}, \bibinfo{author}{T.~D. Noia},
\newblock \bibinfo{title}{Datarec: {A} python library for standardized and reproducible data management in recommender systems},
\newblock in: \bibinfo{booktitle}{Proceedings of the 48th International {ACM} {SIGIR} Conference on Research and Development in Information Retrieval, {SIGIR} 2025, Padua, Italy July 13-18, 2025}, \bibinfo{publisher}{{ACM}}, \bibinfo{year}{2025}. \URLprefix \url{https://doi.org/10.1145/3726302.3730320}. \DOIprefix\doi{10.1145/3726302.3730320}.
\bibitem[{Paparella et~al.(2023)Paparella, Anelli, Nardini, Perego, and Noia}]{DBLP:conf/cikm/PaparellaAN0N23}
\bibinfo{author}{V.~Paparella}, \bibinfo{author}{V.~W. Anelli}, \bibinfo{author}{F.~M. Nardini}, \bibinfo{author}{R.~Perego}, \bibinfo{author}{T.~D. Noia},
\newblock \bibinfo{title}{Post-hoc selection of pareto-optimal solutions in search and recommendation},
\newblock in: \bibinfo{booktitle}{{CIKM}}, \bibinfo{publisher}{{ACM}}, \bibinfo{year}{2023}, pp. \bibinfo{pages}{2013--2023}.
\bibitem[{Anelli et~al.(2021)Anelli, Bellog{\'{\i}}n, Ferrara, Malitesta, Merra, Pomo, Donini, and Noia}]{DBLP:conf/sigir/AnelliBFMMPDN21}
\bibinfo{author}{V.~W. Anelli}, \bibinfo{author}{A.~Bellog{\'{\i}}n}, \bibinfo{author}{A.~Ferrara}, \bibinfo{author}{D.~Malitesta}, \bibinfo{author}{F.~A. Merra}, \bibinfo{author}{C.~Pomo}, \bibinfo{author}{F.~M. Donini}, \bibinfo{author}{T.~D. Noia},
\newblock \bibinfo{title}{Elliot: {A} comprehensive and rigorous framework for reproducible recommender systems evaluation},
\newblock in: \bibinfo{booktitle}{{SIGIR}}, \bibinfo{publisher}{{ACM}}, \bibinfo{address}{New York, NY, USA}, \bibinfo{year}{2021}, pp. \bibinfo{pages}{2405--2414}.
\bibitem[{Ferrara et~al.(2023)Ferrara, Anelli, Mancino, Noia, and Sciascio}]{10.1145/3588901}
\bibinfo{author}{A.~Ferrara}, \bibinfo{author}{V.~W. Anelli}, \bibinfo{author}{A.~C.~M. Mancino}, \bibinfo{author}{T.~D. Noia}, \bibinfo{author}{E.~D. Sciascio},
\newblock \bibinfo{title}{Kgflex: Efficient recommendation with sparse feature factorization and knowledge graphs},
\newblock \bibinfo{journal}{ACM Trans. Recomm. Syst.}  (\bibinfo{year}{2023}).
\bibitem[{Javed et~al.(2021)Javed, Shaukat, Hameed, Iqbal, Alam, and Luo}]{javed2021review}
\bibinfo{author}{U.~Javed}, \bibinfo{author}{K.~Shaukat}, \bibinfo{author}{I.~A. Hameed}, \bibinfo{author}{F.~Iqbal}, \bibinfo{author}{T.~M. Alam}, \bibinfo{author}{S.~Luo},
\newblock \bibinfo{title}{A review of content-based and context-based recommendation systems},
\newblock \bibinfo{journal}{International Journal of Emerging Technologies in Learning (iJET)} \bibinfo{volume}{16} (\bibinfo{year}{2021}) \bibinfo{pages}{274--306}.
\bibitem[{Paudel et~al.(2017)Paudel, Christoffel, Newell, and Bernstein}]{DBLP:journals/tiis/PaudelCNB17}
\bibinfo{author}{B.~Paudel}, \bibinfo{author}{F.~Christoffel}, \bibinfo{author}{C.~Newell}, \bibinfo{author}{A.~Bernstein},
\newblock \bibinfo{title}{Updatable, accurate, diverse, and scalable recommendations for interactive applications},
\newblock \bibinfo{journal}{{ACM} Trans. Interact. Intell. Syst.} \bibinfo{volume}{7} (\bibinfo{year}{2017}) \bibinfo{pages}{1:1--1:34}.
\bibitem[{He et~al.(2020)He, Deng, Wang, Li, Zhang, and Wang}]{DBLP:conf/sigir/0001DWLZ020}
\bibinfo{author}{X.~He}, \bibinfo{author}{K.~Deng}, \bibinfo{author}{X.~Wang}, \bibinfo{author}{Y.~Li}, \bibinfo{author}{Y.~Zhang}, \bibinfo{author}{M.~Wang},
\newblock \bibinfo{title}{Lightgcn: Simplifying and powering graph convolution network for recommendation},
\newblock in: \bibinfo{booktitle}{{SIGIR}}, \bibinfo{year}{2020}, pp. \bibinfo{pages}{639--648}.
\bibitem[{Cooper et~al.(2014)Cooper, Lee, Radzik, and Siantos}]{DBLP:conf/www/CooperLRS14}
\bibinfo{author}{C.~Cooper}, \bibinfo{author}{S.~Lee}, \bibinfo{author}{T.~Radzik}, \bibinfo{author}{Y.~Siantos},
\newblock \bibinfo{title}{Random walks in recommender systems: exact computation and simulations},
\newblock in: \bibinfo{booktitle}{{WWW} (Companion Volume)}, \bibinfo{year}{2014}, pp. \bibinfo{pages}{811--816}.
\bibitem[{Sarwar et~al.(2000)Sarwar, Karypis, Konstan, and Riedl}]{DBLP:conf/sigecom/SarwarKKR00}
\bibinfo{author}{B.~M. Sarwar}, \bibinfo{author}{G.~Karypis}, \bibinfo{author}{J.~A. Konstan}, \bibinfo{author}{J.~Riedl},
\newblock \bibinfo{title}{Analysis of recommendation algorithms for e-commerce},
\newblock in: \bibinfo{booktitle}{{EC}}, \bibinfo{publisher}{{ACM}}, \bibinfo{address}{New York, NY, USA}, \bibinfo{year}{2000}, pp. \bibinfo{pages}{158--167}.
\bibitem[{Breese et~al.(1998)Breese, Heckerman, and Kadie}]{DBLP:conf/uai/BreeseHK98}
\bibinfo{author}{J.~S. Breese}, \bibinfo{author}{D.~Heckerman}, \bibinfo{author}{C.~M. Kadie},
\newblock \bibinfo{title}{Empirical analysis of predictive algorithms for collaborative filtering},
\newblock in: \bibinfo{booktitle}{{UAI}}, \bibinfo{year}{1998}, pp. \bibinfo{pages}{43--52}.
\bibitem[{Steck(2019)}]{DBLP:conf/www/Steck19}
\bibinfo{author}{H.~Steck},
\newblock \bibinfo{title}{Embarrassingly shallow autoencoders for sparse data},
\newblock in: \bibinfo{booktitle}{{WWW}}, \bibinfo{publisher}{{ACM}}, \bibinfo{address}{New York, NY, USA}, \bibinfo{year}{2019}, pp. \bibinfo{pages}{3251--3257}.
\bibitem[{Rendle et~al.(2020)Rendle, Krichene, Zhang, and Anderson}]{DBLP:conf/recsys/RendleKZA20}
\bibinfo{author}{S.~Rendle}, \bibinfo{author}{W.~Krichene}, \bibinfo{author}{L.~Zhang}, \bibinfo{author}{J.~R. Anderson},
\newblock \bibinfo{title}{Neural collaborative filtering vs. matrix factorization revisited},
\newblock in: \bibinfo{booktitle}{RecSys}, \bibinfo{year}{2020}, pp. \bibinfo{pages}{240--248}.
\bibitem[{He et~al.(2017)He, Liao, Zhang, Nie, Hu, and Chua}]{DBLP:conf/www/HeLZNHC17}
\bibinfo{author}{X.~He}, \bibinfo{author}{L.~Liao}, \bibinfo{author}{H.~Zhang}, \bibinfo{author}{L.~Nie}, \bibinfo{author}{X.~Hu}, \bibinfo{author}{T.~Chua},
\newblock \bibinfo{title}{Neural collaborative filtering},
\newblock in: \bibinfo{booktitle}{{WWW}}, \bibinfo{year}{2017}, pp. \bibinfo{pages}{173--182}.
\bibitem[{Salton et~al.(1975)Salton, Wong, and Yang}]{DBLP:journals/cacm/SaltonWY75}
\bibinfo{author}{G.~Salton}, \bibinfo{author}{A.~Wong}, \bibinfo{author}{C.~Yang},
\newblock \bibinfo{title}{A vector space model for automatic indexing},
\newblock \bibinfo{journal}{Commun. {ACM}} \bibinfo{volume}{18} (\bibinfo{year}{1975}) \bibinfo{pages}{613--620}.
\bibitem[{Gantner et~al.(2011)Gantner, Rendle, Freudenthaler, and Schmidt{-}Thieme}]{DBLP:conf/recsys/GantnerRFS11}
\bibinfo{author}{Z.~Gantner}, \bibinfo{author}{S.~Rendle}, \bibinfo{author}{C.~Freudenthaler}, \bibinfo{author}{L.~Schmidt{-}Thieme},
\newblock \bibinfo{title}{Mymedialite: a free recommender system library},
\newblock in: \bibinfo{booktitle}{RecSys}, \bibinfo{publisher}{{ACM}}, \bibinfo{address}{New York, NY, USA}, \bibinfo{year}{2011}, pp. \bibinfo{pages}{305--308}.
\bibitem[{Chen et~al.(2023)Chen, Fu, Yuan, Wen, Fan, Liu, Zhang, Li, and Xiao}]{DBLP:conf/cikm/ChenFYWFL0LX23}
\bibinfo{author}{Y.~Chen}, \bibinfo{author}{Q.~Fu}, \bibinfo{author}{Y.~Yuan}, \bibinfo{author}{Z.~Wen}, \bibinfo{author}{G.~Fan}, \bibinfo{author}{D.~Liu}, \bibinfo{author}{D.~Zhang}, \bibinfo{author}{Z.~Li}, \bibinfo{author}{Y.~Xiao},
\newblock \bibinfo{title}{Hallucination detection: Robustly discerning reliable answers in large language models},
\newblock in: \bibinfo{booktitle}{{CIKM}}, \bibinfo{year}{2023}, pp. \bibinfo{pages}{245--255}.
\bibitem[{Nie et~al.(2019)Nie, Yao, Wang, Pan, and Lin}]{DBLP:conf/acl/NieYWPL19}
\bibinfo{author}{F.~Nie}, \bibinfo{author}{J.~Yao}, \bibinfo{author}{J.~Wang}, \bibinfo{author}{R.~Pan}, \bibinfo{author}{C.~Lin},
\newblock \bibinfo{title}{A simple recipe towards reducing hallucination in neural surface realisation},
\newblock in: \bibinfo{booktitle}{{ACL} {(1)}}, \bibinfo{year}{2019}, pp. \bibinfo{pages}{2673--2679}.
\bibitem[{Ji et~al.(2023)Ji, Lee, Frieske, Yu, Su, Xu, Ishii, Bang, Madotto, and Fung}]{DBLP:journals/csur/JiLFYSXIBMF23}
\bibinfo{author}{Z.~Ji}, \bibinfo{author}{N.~Lee}, \bibinfo{author}{R.~Frieske}, \bibinfo{author}{T.~Yu}, \bibinfo{author}{D.~Su}, \bibinfo{author}{Y.~Xu}, \bibinfo{author}{E.~Ishii}, \bibinfo{author}{Y.~Bang}, \bibinfo{author}{A.~Madotto}, \bibinfo{author}{P.~Fung},
\newblock \bibinfo{title}{Survey of hallucination in natural language generation},
\newblock \bibinfo{journal}{{ACM} Comput. Surv.} \bibinfo{volume}{55} (\bibinfo{year}{2023}) \bibinfo{pages}{248:1--248:38}.
\bibitem[{Giuliano et~al.(1961)Giuliano, Jr., Kimball, Meyer, and Stein}]{DBLP:journals/iandc/GiulianoJKMS61}
\bibinfo{author}{V.~E. Giuliano}, \bibinfo{author}{P.~E.~J. Jr.}, \bibinfo{author}{G.~E. Kimball}, \bibinfo{author}{R.~F. Meyer}, \bibinfo{author}{B.~A. Stein},
\newblock \bibinfo{title}{Automatic pattern recognition by a gestalt method},
\newblock \bibinfo{journal}{Inf. Control.} \bibinfo{volume}{4} (\bibinfo{year}{1961}) \bibinfo{pages}{332--345}.
\bibitem[{Petrov and MacDonald(2023)}]{DBLP:conf/recsys/PetrovM23}
\bibinfo{author}{A.~V. Petrov}, \bibinfo{author}{C.~MacDonald},
\newblock \bibinfo{title}{gsasrec: Reducing overconfidence in sequential recommendation trained with negative sampling},
\newblock in: \bibinfo{booktitle}{RecSys}, \bibinfo{year}{2023}, pp. \bibinfo{pages}{116--128}.
\bibitem[{Olson and Delen(2008)}]{DBLP:books/daglib/0022145}
\bibinfo{author}{D.~L. Olson}, \bibinfo{author}{D.~Delen}, \bibinfo{title}{Advanced Data Mining Techniques}, \bibinfo{publisher}{Springer}, \bibinfo{address}{{US}}, \bibinfo{year}{2008}.
\bibitem[{J{\"{a}}rvelin and Kek{\"{a}}l{\"{a}}inen(2002)}]{DBLP:journals/tois/JarvelinK02}
\bibinfo{author}{K.~J{\"{a}}rvelin}, \bibinfo{author}{J.~Kek{\"{a}}l{\"{a}}inen},
\newblock \bibinfo{title}{Cumulated gain-based evaluation of {IR} techniques},
\newblock \bibinfo{journal}{{ACM} Trans. Inf. Syst.} \bibinfo{volume}{20} (\bibinfo{year}{2002}) \bibinfo{pages}{422--446}.
\bibitem[{Di~Palma et~al.(2025)Di~Palma, Merra, Sfilio, Anelli, Narducci, and Di~Noia}]{di2025llms}
\bibinfo{author}{D.~Di~Palma}, \bibinfo{author}{F.~A. Merra}, \bibinfo{author}{M.~Sfilio}, \bibinfo{author}{V.~W. Anelli}, \bibinfo{author}{F.~Narducci}, \bibinfo{author}{T.~Di~Noia},
\newblock \bibinfo{title}{Do llms memorize recommendation datasets? a preliminary study on movielens-1m},
\newblock in: \bibinfo{booktitle}{Proceedings of the 48th International {ACM} {SIGIR} Conference on Research and Development in Information Retrieval, {SIGIR} 2025, Padua, Italy July 13-18, 2025}, \bibinfo{publisher}{{ACM}}, \bibinfo{year}{2025}. \URLprefix \url{https://doi.org/10.1145/3726302.3730178}. \DOIprefix\doi{10.1145/3726302.3730178}.

\end{thebibliography}


\end{document}